\documentstyle[12pt]{article}
\input{amssym.def}
\input{amssym.tex}
\newcommand{\beq}[1]{\begin{equation}\label{#1}}
\newcommand{\eeq}{\end{equation}}
\newcommand{\bear}[1]{\begin{eqnarray}\label{#1}}
\newcommand{\ear}{\end{eqnarray}}
\newcommand{\bearr}[1]{\begin{eqnarray}\label{#1}\lal}

\newcommand{\nn}{\nonumber}
\def\nnv{\nonumber\\[5pt] {}}

\def\nnnv{\nonumber\\[5pt] \lal }

\newcommand{\rf}[1]{(\ref{#1})}

\def\nhq{\hspace{-0.5em}}
\def\nqq{\hspace{-2em}}
\def\al{&\nhq}
\def\lal{&&\nqq}               
\def\eql{\al=\al}

\def\cm{\hspace{1cm}}
\def\inch{\hspace{1in}}

\def\yy{\\[5pt]}
\def\yyy{\\[5pt] \lal}

\def\eq{Eq.\,}
\def\eqs{Eqs.\,}
\def\eq{Eq.\,}
\def\eqs{Eqs.\,}

\def\e{{\rm e}}
\def\m{{\rm m}}
\def\E{{\rm E}}
\def\M{{\rm M}}
\def\EQ{{\rm EQ}}
\def\MQ{{\rm MQ}}
\def\Qie{Q_{\e I}}
\def\Qim{Q_{\m I}}

\def\half{{\frac{1}{2}}}

\newcommand{\vol}{\mbox{\rm vol}}

\newcommand{\partlx}[1]{\frac{\partial}{\partial x^{#1}}}
\newcommand{\partltau}[1]{\frac{\partial}{\partial \tau^{#1}}}
\newcommand{\p}{\partial}

\newcommand{\btu}{\bigtriangleup}
\newcommand{\sq}[1]{\sqrt{|#1|}}
\newcommand{\ints}{ \mbox{\rm int} }
\def\bh{black hole}
\def\bhs{black holes}%
\newcommand\oG{\overline{G}}
\newcommand{\ol}{ \overline}
\newcommand{\oc}{ \overline{c}}
\newcommand\oI{\overline{I}}
\newcommand\eqdef{\stackrel{\rm def}{=}}

\newcommand{\nl}{ {\hfill \break} }
\newcommand{\np}{ {\newpage } }
\renewcommand{\ol}{ \overline}

\newcommand{\inc}{ {\hookrightarrow } }
\newcommand{\bdy}{ {\partial } }

\newcommand{\Ric}{ {\rm Ric} }

\newcommand{\Diff}{ {\rm Diff} }
\newcommand{\Geom}{ {\rm Geom} }
\newcommand{\Isom}{{\rm Isom} }
\newcommand{\Met}{ {\rm Met} }

\newcommand{\GHY}{ {\rm GHY} }

\newcommand{\BD}{ {\rm BD} }

\newcommand{\so}{ {\rm so} }
\newcommand{\su}{ {\rm su} }

\newcommand{\diag}{ {\rm diag} }

\newcommand{\sign}{ {\rm sign} }
\newcommand{\pr}{ {\rm pr} }
\newcommand{\eps}{ \varepsilon }

\newcommand\mustbe{\stackrel{!}{=}}
\newcommand\mustbeleq{\stackrel{!}{\leq}}
\newcommand{\tr}{\mbox{\rm Tr}\,}

\newcommand{\Cinf}{C^{\infty}}

\def\dst{\displaystyle}
\def\Half{{\dst\frac{1}{2}}}

\def\const{{\rm const}}
\def\DAL{\raisebox{-1.6pt}{\large $\Box$}\,}
\newcommand{\vars}[1]{\left\{\begin{array}{ll}#1\end{array}\right.}
\def\mO{{[-1]\mathstrut}}
\def\pO{{[1]\mathstrut}}
\def\rank{\mathop{\rm rank}\nolimits}

\def\eop{\hfill\mbox{$\Box$}\break}

\newcommand{\T}{{\frak T}}
\newcommand{\N}{ \mbox{\rm I$\!$N} }
\newcommand{\R}{ \mbox{\rm I$\!$R} }
\def\C{\mbox{\rm {I\kern-.520em C}}}
\begin{document}
%
\title{
\vspace*{-1.0cm}
The effective $\sigma$-model of multidimensional gravity}
\author{Martin Rainer\thanks{E-mail: rainer@phys.psu.edu}\\
Center for Gravitational Physics and Geometry,
\\
104 Davey Laboratory, The Pennsylvania State University,
\\
University Park, PA 16802-6300, USA
\\
and
\\
Gravitationsprojekt, Mathematische Physik I,
\\
Institut f\"ur Mathematik,Universit\"at Potsdam,
\\
PF 601553, D-14415 Potsdam, Germany
}

\date{October 1998}

\maketitle
\vspace*{-9cm}
\vspace*{9cm}

\begin{abstract}
The properties of the
effective $\sigma$-model for $D$-dimensional Einstein gravity
based on multidimensional geometries is analyzed.
Besides pure geometry additional
minimally coupled scalars and $(p+2)$-forms
are considered which yield an extended target space
after reduction to the effective $D_0$-dimensional geometry.

The target space is always a homogeneous space. Exact solutions
exist provided an orthobrane condition is satisfied which
geometrically makes the target space a locally symmetric one.

New solutions with scalar fields are found
which may inflate not only in time-like but in also in additional
spatial directions of the effective geometry.

Static spherically symmetric solutions with a particular
configuration of intersecting electric and magnetic branes are
investigated both, for the orthobrane case and for degenerated
charges. In both cases  $T_H$ depends critically on the
intersection dimension of the branes.

Finally, the role of the Einstein frame for $4$-geometries is addressed, and
the physical frame transformation for cosmological geometries is given.
\end{abstract}
%
\nl
\section{\bf Introduction}
\setcounter{equation}{0}
Historically $\sigma$-models have turned out to be a very powerful
tool in many areas of physics. In gravity the importance was soon
realized \cite{BMG} in the context of solution generating
techniques \cite{Ger}. More recently, $\sigma$-models have been
also discussed in the context of string theory \cite{Gal,GalR,KS}.

The purpose of this paper is to clarify the geometric structure of
the effective $\sigma$-model for multidimensional Einstein
geometry and to demonstrate its applicability in such different
directions as cosmology, (extended) string theory, and
quantization of certain higher-dimensional geometric actions.

In fact it turns out to be a very powerful tool which, on one
side, allows to test the geometric content of string and M-theory
down to their concrete physical imprints in the physical
space-time and, on the other side, prepares a well defined class
of classical higher-dimensional geometries  for the canonical
quantization program in dimension $D_0\leq 4$ whenever this is
applicable to pure Einstein gravity itself. In principle all cases
with infinite number of degrees of freedom in dimension $D_0=4$
which can be canonically quantized have some analogous cases where
additional extra dimensions add only a finite number of degrees of
freedom without disturbing the integrability of the problem. These
cases include of course also recently investigated midisuperspace
$4$-geometries. In the case of spherical symmetries, and more
particular in the static case, one can find particular solutions
to a classical system of the multidimensional Einstein action with
scalar and antisymmetric $p+2$-form fields which are
multidimensional extensions of black hole solutions. It turns out
that the standard surface gravity and the Hawking temperature
$T_H$ as calculated from a Komar-like integral depend sensitively
on the intersection dimension of the $p$-branes involved in the
solution. This provides, at least in principle, an observational
window to very direct geometrical properties of possible extra
dimensions. Apart from that, the multidimensional $\sigma$-model
contains all kinds of multidimensional spatially homogeneous
cosmological models as degenerate minisuperspace cases with a
finite number of degrees of freedom only.

Below, the effective $D_0$-dimensional $\sigma$-model is derived
from a multidimensional action of Einstein type in a higher
dimension $D$, first for pure geometry, then with additional
scalar and antisymmetric $p+2$-form matter fields. The domains of
the $p+1$-form potentials of the antisymmetric $p+2$-forms are the
world-sheets of $p$-branes. In extended string and M-theory
\cite{SV,Du,SchSch} strings are generalized to membranes as
higher-dimensional objects. Most of these unified models are
modeled initially on a higher-dimensional space-time manifold, say
of dimension $D>4$, which then undergoes some scheme of
spontaneous compactification.

The geometric structure of the target-space is clarified. In
particular it is shown that it is a always a homogeneous space. It
is furthermore locally symmetric if and only if the characteristic
target-space vectors satisfy a particular orthogonality condition,
called the {\em orthobrane} relation whenever they are not
identical. In any case, it turns out possible to express the {\em
general} exact solutions in terms of elementary functions,
provided the input parameters of the model satisfy the , whence
the target space is locally symmetric.

Solutions of the corresponding field equations are discussed
generally and with concrete examples.
Particular solutions for the subcases with
Ricci flat internal spaces with scalar fields only,
and with intersecting p-branes are presented.
In the subcase of spherically symmetric solutions
the relation to particles and black $p$-branes is given.
Although a priori one might admit all possible types of
components of $F$-fields compatible with spherical symmetry,
namely, electric, magnetic and quasiscalar ones,
we concentrate on true electric and magnetic type fields,
since these are the ones which admit black hole solutions.

Besides the orthobrane solutions which by now became popular in
string theory, there are further families of solutions, which have
another additional symmetry, e.g.  coinciding $F$-field charges
for the electro-magnetic solutions. In target space this
additional symmetry is expressed by a linear relation between
certain column vectors  of the coupling matrix. In this case the
original orthobrane conditions reduce to some weaker set of
orthogonality conditions.

In the case of static, spherical symmetric solutions it is
demonstrated that the formal Hawking temperature $T_H$ (as it
might appear to an observer at infinity) depends sensitively on
the intersection dimension of the $p$-branes. Hence solutions to
the multidimensional $\sigma$-model allow to detect possible
imprints from extra-dimensional internal factor spaces within the
physical dimension $D_0=4$. The \bh\ solutions depend on 3
integration constants, related to the electric, the magnetic, and
the mass charge. It is also shown that the Hawking temperature of
such \bhs\ depends on the intersection dimension $d_{\ints}$ of
the corresponding $p$-branes. In an extremal limit of the charges,
the \bh\ temperature turns out to converge to zero for
$d_{\ints}=0$, to a finite limit for $d_{\ints}=1$, and to
infinity for $d_{\ints} > 1$.

Finally it is shown how the geometries of well known solutions in
a Brans-Dicke frame can be transformed to the physically relevant
Einstein frame.
\section{\bf Pure multidimensional gravity}
\label{Sect. 2}
\setcounter{equation}{0}
For the purpose of this paper let a
($\Cinf$-) {\em multidimensional} (MD) manifold $N$
be topologically just defined by
a $\Cinf$-fiber bundle
\beq{1.1}
M \inc N \rightarrow \overline{M}_{0}
\eeq
with a direct product
\beq{1.x0}
M:=\times_{i=1}^{n} M_{i}
\eeq
of internal $\Cinf$ factor spaces $M_i$, $i=1,\ldots,n$,
as a standard fiber,
and a distinguished  $\Cinf$ base manifold $\overline{M}_{0}$.
(Later, for considerations of dynamics and cosmology
we will set in particular $\overline{M}_{0}:=\R\times M_0$,
and for the connection representation of Einstein gravity
${D}_0:=4$ will be required.)

%
The MD manifold $N$ is called {\em internally homogeneous}
if there exists a direct product group
$G:=\bigotimes_{i=0}^{n} G_{i}$
with a direct product realization $\tau:=\otimes_{i=0}^{n} \tau_i$
on $\Diff(M):=\bigotimes_{i=0}^{n}\Diff(M_i)$ such that
for $i=0,\ldots,n$ the realization
\beq{1.x1}
\tau_i:G_{i}\to \Diff(M_i)
\eeq
yields a transitive action of $\tau_i(G_{i})$ on $M_i$.

{\bf Definition:}
A ($\Cinf$) Riemannian manifold $(M,g)$ (of arbitrary signature)
is a $\Cinf$ manifold $M$ equipped with
a symmetric bilinear $\Cinf$ section $g:M \to \frak T^0_2 M$
called metric.
Unless specified otherwise the metric $g$ will
always be assumed to be non-degenerate.
\hfill\mbox{$\Box$}\break

{\bf Definition:}
Given a Riemannian manifold $(M,g)$,
a diffeomorphism $\chi\in\Diff(M)$
is called an {\em isometry} of $(M,g)$
whenever it leaves $g$ invariant,
i.e. whenever
\beq{1.x2}
g_{\chi (p)}=g_{p} \quad \forall p\in M .
\eeq
\eop

The very fact that a given diffeomorphism $\chi\in \Diff(M)$
may be an isometry on some metric but not on another one
is the reason why the action of $\Diff(M)$ is not free on the
space $\Met(M)$ of $\Cinf$-metrics on $M$,
whence $\Geom(M):=\Met(M)/\Diff(M)$ is in general not a manifold.

{\bf Definition:} A Riemannian manifold $(M,g)$ is called
homogeneous, whenever $M$ is homogeneous with a corresponding
group $G$ having a transitive realization $\tau(G)\subset\Diff(M)$
which leaves $g$ invariant, i.e. \beq{1.x3} g_{\chi
(p)}=g_{p}\quad \forall p\in M\quad \forall\chi\in \tau(G) . \eeq
\eop

Now for $i=0,\ldots,n$, let each factor space $M_i$ be equipped
with a smooth homogeneous metric $g^{(i)}$. rendering it into a
homogeneous Riemannian manifold. Furthermore, let
$\overline{M}_{0}$ be equipped with an arbitrary $\Cinf$-metric
$\overline{g}^{(0)}$, and let $\ol\gamma$ and $\beta^i$ ,
$i=1,\ldots,n$ be smooth scalar fields on $\overline{M}_{0}$.

Then,
under any projection $\pr: N\to \overline{M}_{0}$
a pullback
of $e^{2\ol\gamma}\overline{g}^{(0)}$
from $x\in \overline{M_{0}}$ to $z\in\pr^{-1}\{x\}\subset M$,
consistent with the fiber bundle \rf{1.1}
and the homogeneity of internal spaces,
is given by
\beq{1.2}
g_{(z)}:=e^{2\ol\gamma(x)}\overline{g}^{(0)}_{(x)}
\oplus_{i=1}^{n} e^{2\beta^i(x)}g^{(i)} .
\eeq
The function $\ol\gamma$
fixes a {\em gauge} for the (Weyl)
{\em conformal frame} on $\overline{M_0}$,
corresponding just to a particular choice
of geometrical variables.

$\ol\gamma$ uniquely defines
the form of the effective ${D}_0$-dimensional theory.
For example $\ol\gamma:=0$ defines the
Brans-Dicke frame,

Let us now consider a multidimensional manifold $N$
\rf{1.1} of dimension $D={D}_0 + \sum_{i=0}^{n} d_i $,
equipped with
a (pseudo) Riemannian metric
\rf{1.2} where
\beq{2.1}
g^{(i)} \equiv g_{m_{i} n_{i}}(y_i) dy_i^{m_{i}} \otimes dy_i^{n_{i}} ,
\eeq
are  $R$-homogeneous Riemannian metrics on $M_i$
(i.e. the Ricci scalar $R[g^{(i)}] \equiv R_i$ is a constant on $M_i$),
in coordinates $y_i^{n_{i}}$,
$n_{i}=0,\ldots, d_{i}$,
and
\beq{2.2}
x\mapsto \ol{g}^{(0)}(x)
=\ol{g}^{(0)}_{\mu\nu}(x) d x^{\mu} \otimes d x^{\nu}
\eeq
yielding a general, not necessarily $R$-homogeneous,
(pseudo) Riemannian metric on $\ol{M}_0$.

With \rf{1.2} is a multidimensional generalization of
the warped product of \cite{BeemEE},
namely $N=\ol{M}_0\times_a M$, where $a:=e^\beta$ is
now a {\em vector-valued} root warping function, given
by
\beq{betavec}
\beta:=\left(
\begin{array}{l}
\beta_0
\\
\vdots
\\
\beta_n
\end{array}
\right) .
\eeq
Below sometimes, in particular for physical application to
the $\Diff(\ol M_0)$-invariant case with ${D}_0=4$,
we will assume the $i=0$ geometry to be empty and omit
corresponding empty contributions to tensors, summations
etc.
For later convenience we also define
\beq{e14}
         \eps (I) := \prod_{i\in I} \eps_i  ;       \cm
         \sigma_0 := \sum_{i=0}^{n} d_i \beta_i  ,  \cm
         \sigma_1 := \sum_{i=1}^{n} d_i \beta_i  ,  \cm
        \sigma(I):= \sum_{i\in I} d_i \beta_i  ,
\eeq where $\eps_i:=\sign(|g^{(i)}|)$ and $M_i\subset M$ for
$i=0,\ldots,n$ are all homogeneous factor spaces. Here and below,
we use the shorthand $|g| := |\det (g_{MN})|$, $|\ol{g}^{(0)}| :=
|\det (\ol{g}^{(0)}_{\mu\nu})|$, and analogously for all other
metrics including $g^{(i)}$, $i=1, \ldots, n$.

Further,
a $\overline{g}^{(0)}$-covariant derivative of a given function $\alpha$
w.r.t. $x^\mu$ is denoted  by  $\alpha_{;\mu}$,
its partial derivative also by $\alpha_{,\mu}$,
and $(\p\alpha)(\p\beta):=\ol{g}^{(0)\mu\nu} \alpha_{,\mu} \beta_{,\nu} $.

On $\overline{M}_0$, the Laplace-Beltrami operator
$\Delta[\overline{g}^{(0)}]=
\frac{1}{\sqrt{|\overline{g}^{(0)}|}}
{\partlx{\mu}}
\left( \sqrt{|\overline{g}^{(0)}|} \overline{g}^{(0)\mu\nu}
{\partlx{\nu}}
\right)
$ ,
transforms under the conformal map
$\overline{g}^{(0)}\mapsto e^{2\overline{\gamma}} \overline{g}^{(0)}$
according to
\bear{2.3}
\Delta[e^{2\overline{\gamma}}\overline{g}^{(0)}]&=&
e^{-2\overline{\gamma}}\Delta[\overline{g}^{(0)}]
-e^{-2\overline{\gamma}}
{\overline{g}^{(0)}}^{\mu\nu}
\left(
\Gamma[e^{2\overline{\gamma}}\overline{g}^{(0)}]-\Gamma[\overline{g}^{(0)}]
\right)^{\lambda}_{\mu\nu}
{\partlx{\lambda}}
\nn\\
&=&
e^{-2\ol{\gamma}}
\left(
\Delta[\ol{g}^{(0)}]
+({D}_0-2){g^{(0)}}^{\mu\nu}
\frac{\partial \ol\gamma}{\partial x^\mu}
{\partlx{\nu}}
\right) ,
\ear
where $\Gamma$ denotes the Levi-Civita connection.

The Levi-Civita connection $\Gamma$ corresponding to \rf{1.2}
does {\em not} decompose multidimensionally, and
neither does the Riemann tensor.
The latter is a section in $\T^1_3M$
which is not given as a pullback to $\ol M_0$ of a section in
the direct sum $\oplus_{i=1}^{n} \T^1_3M_i$
of  corresponding tensor bundles over the factor manifolds.

However, with (\ref{1.2})
the Ricci tensor decomposes again multidimensionally:
\beq{1.2R}
\Ric[g]=
\Ric^{(0)}[g^{(0)}, {\ol\gamma}
; \phi
]
\oplus^{n}_{i=1}
\Ric^{(i)}[g^{(0)}, {\ol\gamma}%
; g^{(i)},\phi
] ,
\eeq
where
\bear{2.25}
\Ric^{(0)}_{\mu \nu}
& :=  &
R_{\mu \nu}[g^{(0)} ]
+ g^{(0)}_{\mu \nu} \Bigl\{
- \btu[g^{(0)}] {\ol\gamma} +  (2-D_0)  (\p {\ol\gamma})^2
- \p {\ol\gamma} \sum_{j=1}^{n} d_j \p \phi^j
\Bigr\}
\nn
\\
&  &
+ (2 - D_0) ({\ol\gamma}_{;\mu \nu} - {\ol\gamma}_{,\mu} {\ol\gamma}_{,\nu})
- \sum_{i=1}^{n} d_i ( \phi^i_{;\mu \nu} - \phi^i_{,\mu} {\ol\gamma}_{,\nu}
- \phi^i_{,\nu} {\ol\gamma}_{,\mu} + \phi^i_{,\mu} \phi^i_{,\nu}) ,
\nn\\
\label{2.26}
\Ric^{(i)}_{m_{i} n_{i}}
&:= &
{R_{m_{i}n_{i}}} [g^{(i)}]
- e^{2 \phi^{i} - 2 {\ol\gamma}} g^{(i)}_{m_{i}n_{i}}
      \biggl\{ \btu[g^{(0)}] \phi^{i}
+ (\p \phi^{i}) [ (D_0 - 2) \p {\ol\gamma}  +
          \sum_{j=1}^{n} d_j \p \phi^j ] \biggr\} ,
\nn
\\
&  &
i = 1, \ldots, n ,
\ear

The corresponding Ricci curvature scalar
reads
\bear{2.4}
R[g] &=&
e^{-2\ol{\gamma}}R[\ol{g}^{(0)}]
+\sum_{i=1}^{n} e^{-2\beta^i} R[g^{(i)}]
- e^{-2\ol{\gamma} }
\ol{g}^{(0)\mu\nu}
\left(
({D}_0-2)({D}_0-1)
\frac{\partial\ol{\gamma}}{\partial x^\mu}
\frac{\partial\ol{\gamma}}{\partial x^\nu}
\right.
\nn\\
& &
\left.
+ \sum_{i,j=1}^{n} (d_i\delta_{ij}+d_i d_j)
\frac{\partial\beta^i}{\partial x^\mu}
\frac{\partial\beta^j}{\partial x^\nu}
+2({D}_0-2)\sum_{i=1}^{n}d_i
\frac{\partial\ol{\gamma} }{\partial x^\mu}
\frac{\partial\beta^i}{\partial x^\nu}
\right)
\nn\\
& &
- 2 e^{-2\ol{\gamma} }
\Delta[\ol{g}^{(0)}]
\left(
({D}_0-1)\ol{\gamma}
+\sum_{i=1}^{n} d_i\beta^i
\right) .
\ear
Let us now set
\beq{2.5}
f \equiv {f}[\ol{\gamma}, \beta]
:= ({D}_0 - 2) \ol{\gamma} +\sum_{j=1}^{n} d_j  \beta^j ,
\eeq
where $\beta$
is the vector field with the dilatonic scalar fields
$\beta^i$ as components.
(Note that $f$ can be resolved for
$\ol\gamma\equiv \ol\gamma[f,\beta]$
if and only if ${D}_0 \neq 2$.
The singular case $D_0=2$ is discussed in \cite{RZ}.)
Then,
\rf{2.4} can also be written as
\bear{2.6}
 R[g] \! &\!-\! &\!
e^{-2\ol{\gamma} } R[\ol{g}^{(0)}]
-\sum_{i=1}^{n} e^{-2\beta^i} R_i \ =
\\\nn
= \!&\!-\! &\!  e^{-2\ol{\gamma} }
\left\{
\sum_{i=1}^{n} d_i (\p\beta^i)^2
+ (\p f)^2
+(D_0-2) (\p\ol{\gamma})^2
+ 2 \Delta[\ol{g}^{(0)}] (f+\ol{\gamma})
\right\}
\\\nn
=\! &\!-\! &\!  e^{-2\ol{\gamma}}
\left\{
\sum_{i=1}^{n} d_i (\p\beta^i)^2
+ ({D}_0 - 2) (\p \ol{\gamma})^2 - (\p f) \p (f + 2 \ol{\gamma})
+ R_{B} \right\} ,
\\
\label{2.7}
R_B &:=& \frac{1}{\sq {\ol{g}^{(0)}} } e^{-f}
     \p_{\mu} \left[2 e^f \sq {\ol{g}^{(0)} }
     {\ol{g}^{(0)\mu \nu}} \p_{\nu} (f + \ol{\gamma}) \right] ,
\ear
where the last term will yield just a boundary contribution \rf{2.12}
to the action \rf{2.11} below.

Let us assume all $M_{i}$, $i=1,\ldots,n$, to be connected and oriented.
The Riemann-Lebesgue volume form on $M_i$ is denoted by
\beq{2.8}
\tau_i  := \vol(g^{(i)})=\sqrt{|g^{(i)}(y_i)|}
\ dy_i^{1} \wedge \ldots \wedge dy_i^{d_i} ,
\eeq
and the total internal space volume by
\beq{2.9}
\mu:= \prod_{i=1}^n \mu_i, \quad
\mu_i := \int_{M_i} \tau_i =\int_{M_i} \vol(g^{(i)})  .
\eeq
If all of  the spaces $M_i$, $i=1,\ldots,n$ are compact,
then the volumes $\mu_i$ and $\mu$ are finite,
and so are also the numbers
$\rho_i=\int_{M_i}\vol({g^{(i)}}) R[g^{(i)}] $.
However, a non-compact $M_i$ might have infinite volume $\mu_i$  or infinite
$\rho_i$.
Nevertheless, by the $R$-homogeneity of $g^{(i)}$
(in particular satisfied for Einstein spaces),
the ratios
$\frac{\rho_i}{\mu_i}=R[g^{(i)}] $,
$i=1,\ldots,n$, are just finite constants.
In any case, the
$D$-dimensional coupling constant $\kappa$
can be tuned such that,
under the dimensional reduction $\pr: M\to \ol{M}_0$,
\beq{2.10}
\kappa_0:=\kappa \cdot \mu^{-\frac{1}{2}}
\eeq
becomes the ${D}_0$-dimensional physical coupling constant.
If ${D}_0=4$, then ${\kappa_0}^2=8\pi G_N$,
where $G_N$ is the Newton constant.
The limit  $\kappa\to\infty$ for $\mu\to\infty$
is in particular
harmless, if $D$-dimensional gravity is given purely by curvature geometry,
without additional matter fields.
If however this geometry is coupled with finite strength
to additional (matter) fields,
one should indeed better take care to
have all internal spaces $M_i$, $i=1,\ldots,n$
compact.
Often this
can be achieved by factorizing with an
appropriate finite symmetry group.

With the total dimension $D$ , $\kappa^2$ a $D$-dimensional
gravitational constant we consider a purely gravitational
action of the form
\beq{2.11}
S = \frac{1}{2\kappa^2} \int_{N} d^{D}z \sq g
\{ R[g] 
\} + S_{\rm GHY} 
 .
\eeq
Here a (generalized) Gibbons-Hawking-York \cite{GH,Y} type
boundary contribution $S_{\rm GHY}$ to the action is taken
to cancel boundary terms.
Eqs.\rf{2.6} and \rf{2.7} show that $S_{\rm GHY}$ should be taken in the form
\bear{2.12}
S_{\rm GHY} &:=& \frac{1}{2\kappa^{2}} \int_{N} d^{D}z \sq g
     \{ e^{-2\ol \gamma} R_{B} \}
\nn\\
& = &\frac{1}{\kappa^2_0}
\int_{\ol{M}_0}d^{{D}_0}x
\partlx{\lambda}
\left(
e^{f}
\sqrt{|{\ol{g}^{(0)}}|} \ol{g}^{(0)\lambda\nu}
\partlx{\nu} (f+\ol{\gamma})
\right) ,
\ear
which is just a pure boundary term in form of an effective
${D}_0$-dimensional flow through $\p \ol{M}_0$.

After dimensional reduction the action \rf{2.11} reads
\bear{2.15}
S=\frac{1}{2\kappa^2_0}
\int_{\ol{M}_0}d^{{D}_0}x
\sqrt{|{\ol{g}^{(0)}}|}
e^f
\left\{
R[\ol{g}^{(0)}]
+(\p f)(\p[f+2\ol{\gamma}])
-\sum_{i=1}^{n} d_i (\p\beta^i)^2
\right.
\nn\\
\left.
-({D}_0-2) (\p\ol{\gamma})^2
+ e^{2\ol\gamma}\left[\sum_{i=1}^{n}e^{-2\beta^i} R_i
\right]
\right\} ,
\ear
where $e^f$ is a dilatonic scalar field coupling to the
$D_0$-dimensional geometry on
$\ol{M}_0$.

According to the considerations above,
due to the conformal reparametrization invariance
of the geometry on $\ol{M}_0$, we should fix a
conformal frame on $\ol{M}_0$. But then in  \rf{2.15} $\ol{\gamma}$, and
with \rf{2.5} also $f$,
is no longer independent from the vector field $\beta$, but rather
\beq{2.16}
\ol{\gamma}\equiv \ol{\gamma}[\beta]\ ,\qquad f\equiv f[\beta] .
\eeq
Then, modulo the
conformal factor $e^f$,
the dilatonic kinetic term of \rf{2.15}
takes the form
\beq{2.17}
(\p f)(\p[f+2\ol{\gamma}]) -\sum_{i=1}^{n} d_i (\p\beta^i)^2
-({D}_0-2) (\p\ol{\gamma})^2= -  G_{ij} (\p\beta^i) (\p\beta^j) ,\
\eeq
with $G_{ij}\equiv{}^{(\ol{\gamma})}\!G_{ij}$, where
\bear{2.18}
{}^{(\ol{\gamma})}\!G_{ij}&:=& {}^{(\BD)}\!G_{ij}
 -({D}_0-2)({D}_0-1) \, \frac{\p\ol{\gamma}}{\p\beta^i}
\frac{\p\ol{\gamma}}{\p\beta^j}
-2 (D_0-1) d_{(i} \, \frac{\p\ol{\gamma}}{\p\beta^{j)} } ,\
\\
\label{2.19}
&& {}^{(\BD)}\!G_{ij} := \delta_{ij} d_i - d_i d_j .
\ear
For $D_0\neq 2$, we can write equivalently
$G_{ij}\equiv{}^{(f)}\!G_{ij}$, where
\bear{2.20}
{}^{(f)}\!G_{ij} &:=& {}^{(\E)}\!G_{ij}
-\frac{{D}_0-1}{{D}_0-2} \, \frac{\p f}{\p\beta^i}
\frac{\p f}{\p\beta^j} ,\
\\
\label{2.21}
&& {}^{(\E)}\!G_{ij} :=  \delta_{ij} d_i + \frac{d_i d_j}{{D}_0-2} .
\ear
For ${D}_0=1$, $G_{ij}= {}^{(\E)}\!G_{ij}={}^{(\BD)}\!G_{ij}$ is
independent of $\ol{\gamma}$ and $f$.
Note that the metrics \rf{2.19} and \rf{2.21} (with $D_{0}\neq 2$)
may be diagonalized
to $({\mp}({\pm})^{\delta_{1{D}_0}})^{\delta_{1i}} \delta_{ij}$
respectively,
by homogeneous linear minisuperspace coordinate transformations
$\beta {\stackrel{T}{\mapsto}} z$
and $\beta {\stackrel{Q}{\mapsto}} \varphi$,
explicitly given by components
\bear{z}
z^1:={}^{(\BD)}\!q^{-1}\sum_{j=1}^{n}d_j\beta^j\ ,
&\qquad&
\varphi^1:={}^{(\E)}\!q^{-1}\sum_{j=1}^{n}d_j\beta^j\ ,
\nn\\
z^i\equiv\varphi^i&:=&
{\left[\left.d_{i-1}\right/\Sigma_{i-1}\Sigma_{i}\right]}^{1/2}
\sum_{j=i}^{n}
d_j\left(\beta^j-\beta^{i-1}\right)\ ,
\ear
$i=2,\ldots,n$, where
with
${D'}:=D-D_0$ and
$\Sigma_k:=\sum_{i=k}^{n}d_i$,
\bear{qfactor}
{}^{(\BD)}q:=\sqrt{\frac{D'}{D'-1}}\ ,
&\qquad &
{}^{(\E)}q:=\sqrt{\frac{D'(D_0-2)}{D'+D_0-2}}\ .
\ear
So, after fixing a conformal reparametrization gauge for
the geometry on $M_0$,
\rf{2.11} becomes a $\sigma$-model, where
the vector field $\beta$ (or $z$ resp. $\varphi$)
defines the coordinates of its $n$-dimensional target space.
In the following, we will simplify notation by
a summation convention for tensors over target space.

In general, for $n>2$ and non-constant functional $\gamma[\beta]$,
the minisuperspace metric  given by \rf{2.17}
and the conformally related target space metric may
not even be conformally flat. However,
for constant ${\ol\gamma}$, \rf{2.18} reduces to  \rf{2.19},
whence target space is  conformally flat,
namely it is related to $n$-dimensional Minkowski space
by a conformal scale factor
\beq{dil}
\varphi\equiv\varphi(\beta):=\prod_{l=1}^n e^{d_l\beta^l}
=e^{{}^{(\BD)}\!q z^1}=e^{{}^{(\E)}\!q\varphi^1} ,
\eeq
which is proportional to
the total internal space volume.

In the case $D_0\neq 2$, for non-constant functional $f[\beta]$,
the target space may again in general not be conformally flat for
$n>2$. However, for constant $f$, \rf{2.20}
reduces to \rf{2.21}, whence, target space is
a flat $n$-dimensional space, namely
an Euclidean one for $D_0>2$, and a Minkowskian one for $D_0=1$.

After gauging $\ol{\gamma}$, setting $m:=\kappa^{-2}_0$,
\rf{2.15} yields a $\sigma$-model
in the form
\bear{Sgamma0}
{}^{(\ol \gamma)}\!S=
\int_{\ol M_0} d^{{D}_0}x \sqrt{|\ol g^{(0)}|}
{}^{(\ol \gamma)}\!N^{{D}_0} &\!  &\!  \varphi(\beta) \
\left\{
\frac{m}{2} {}^{(\ol \gamma)}\!N^{-2}
\left[ R[\ol g^{(0)}] - {}^{(\ol \gamma)}\!G_{ij}(\p\beta^i)(\p\beta^j)
\right]
\right.
\nn\\
&\! &\! \qquad \qquad
\left.
- {}^{(\BD)}\!V(\beta)
\right\}\ ,
\\
\label{Vgamma0}
 \mbox{\rm where}\qquad
{}^{(\BD)}\!V(\beta) \ :=&\! &\! m\left[
-\frac{1}{2} \sum_{i=1}^{n} R[g^{(i)}]
e^{-2\beta^i} \right] \ ,
\\
\label{Ngamma0}
{}^{(\ol \gamma)}\!N \ :=&\! &\! e^{\ol \gamma}\ .
\ear
Note that, the potential \rf{Vgamma0}
and the conformal factor $\phi(\beta):=\prod\nolimits_{i=1}^{n}e^{d_i\beta^i}$
are gauge invariant.

Analogously, the $\sigma$-model action from \rf{2.15} gauging $f$ can also be
written as
\bear{Sf0}
{}^{(f)}\!S=\int_{\ol M_0} d^{{D}_0}x \sqrt{|\ol g^{(0)}|}
{}^{(f)}\!N^{{D}_0}
 &\! &\!
\left\{\frac{m}{2}
{}^{(f)}\!N^{-2}
\left[ R[\ol g^{(0)}] - {}^{(f)}\!G_{ij}(\p\beta^i)(\p\beta^j)
\right]
\right.
\nn\\
&\! &\! \qquad \qquad
\left.
- {}^{(\E)}\!V(\beta)
\right\} \ ,
\\
\label{Vf0}
{}^{(\E)}\!V(\beta)\ := &\! &\!
m \Omega^2
\left[
-\frac{1}{2} \sum_{i=1}^{n} R[g^{(i)}]
e^{-2\beta^i} \right]\  ,
\\
\label{Nf0}
{}^{(f)}\!N\ := &\! &\! e^{\frac{f}{{D}_0-2}}\ ,
\ear
where the function $\Omega$ on $\ol{M}_0$ is defined as
\beq{Omega}
\Omega:=\varphi^{ \frac{1}{2-{D}_0} }\ .
\eeq
Note that, with $\Omega$ also the potential \rf{Vf0}
is gauge invariant,
and the dilatonic target-space,
though not even conformally flat in general,
is flat for constant $f$.

In fact, Eqs. \rf{Sgamma0}-\rf{Ngamma0} and \rf{Sf0}-\rf{Nf0} show that there
are at least two special frames.

The first one corresponds to the gauge $\ol\gamma \mustbe 0$.
In this case
${}^{(\ol \gamma)}\!N=1$, the minisuperspace metric
\rf{2.18} reduces to the Minkowskian \rf{2.19},
the dilatonic scalar field becomes proportional to the internal space volume,
$e^{f[\beta]}=\varphi(\beta)=\prod\nolimits_{i=1}^{n}e^{d_i\beta^i}$, and
\rf{Sgamma0} describes a generalized $\sigma$-model with conformally
Minkowskian target space.  The Minkowskian signature implies a negative
sign in the dilatonic kinetic term.  This frame is usually called the
Brans-Dicke one, because $\varphi=e^f$ here plays the role of
a Brans-Dicke scalar field.

The second distinguished frame corresponds to the gauge $f \mustbe 0$,
where $\ol\gamma =\frac{1}{2 - D_0}\sum_{i=1}^n d_i\beta^i $
is well-defined only for $D_0\neq 2$.  In this case
${}^{(f)}\!N=1$, the minisuperspace metric
\rf{2.20} reduces to the Euclidean \rf{2.21}, and \rf{Sf0}  describes a
self-gravitating $\sigma$-model with Euclidean target space.  Hence all
dilatonic kinetic terms have positive signs.  This frame is usually called
the Einstein one, because it describes an effective $D_0$-dimensional
Einstein theory with additional minimally coupled scalar fields.
For multidimensional geometries with $D_0 = 2$ the Einstein frame
fails to exist, which reflects the well-known fact that two-dimensional
Einstein equations are trivially satisfied without implying any dynamics.

For ${D}_0=1$, the action of both \rf{Sgamma0} and \rf{Sf0} was shown in
\cite{Ra0} (and previously in \cite{Ra1,Ra2}) to take the form
of a classical particle motion on minisuperspace, whence different frames
correspond are just related by a time reparametrization.
More generally, for ${D}_0\neq 2$ and $(\ol M_0,\ol g{(0)})$
a vacuum space-time,
the $\sigma$-model \rf{Sf0} with the gauge $f \mustbe 0$
describes the dynamics of a massive
$({D}_0-1)$-brane within a potential \rf{Vf0}
on its target minisuperspace.

In fact, the target space is in general a conformally
homogeneous space, and in the Einstein frame a homogeneous one.
Once its isometry group $\frak G$ and isotropy group $\frak H$
are known, it is clear that
the sigma model \rf{Sf0} can also be written in matrix form
\bear{Smatrix}
{}^{(f)}\!S=\int_{\ol M_0} d^{{D}_0}x \sqrt{|\ol g^{(0)}|}
N^{{D}_0}({\cal M})
 &\! &\!
\left\{\frac{m}{2}
N^{-2}({\cal M})
\left[ R[\ol g^{(0)}] + g^{(0)\mu\nu} B
\tr_{\rho} (\p_\mu{\cal M}\p_\nu{\cal M}^{-1})
\right]
\right.
\nn\\
&\! &\! \qquad \qquad
\left.
- {}^{(\E)}\!U({\cal M})
\right\} \ ,
\ear
with ${\cal M}\in\frak \rho(G)$
where
$\rho$
is an appropriate coset representation of the target space
${\frak M}:={\frak G}/{\frak H}$,
${}^{(\E)}\!U$ is now the corresponding potential on $\frak M$,
$N$ a gauge function on $\frak M$, and $B$ a normalization.

For $D_0=4$,
eq. (\ref{Smatrix}) can also be written
in the Einstein frame
as
\bear{Smatq}
{}^{(E)}\!S=\int_{\ol M_0}
 &\! &\!
\left\{\frac{m}{2}
\left[\tr \Omega\wedge *\Sigma + B \tr_{\rho}
d{\cal M}\wedge *d{\cal M}^{-1}
\right]
\right.
\nn\\
&\! &\! \qquad \qquad
\left.
- {}^{(\E)}\!U({\cal M}) *1
\right\} \ ,
\ear
where $\Omega$ is the curvature $2$-form,
$\Sigma:=e\wedge e$ and $\ol g^{(0)}$ are given by
the $D_0$-dimensional soldering $1$-form $e$, and
the Hodge star is taken w.r.t. $({\ol M},g^{(0)})$.
The form \rf{Smatq} is a then a convenient starting
point for the canonical quantization procedure.

In the purely gravitational model consider so far
${\frak M}$ is a finite dimensional
and homogeneous with a transitive Abelian group.
In the following section let us add minimally coupled
scalar and $p+2$-form matter  fields and investigate the extension
of the resulting target space ${\frak M}$.

\section{\bf $\sigma$-model with minimally coupled scalars and $p+2$-forms}
\setcounter{equation}{0}
We now couple the purely gravitational action \rf{2.11}
to additional matter fields of scalar and
generalized Maxwell type, i.e.
we consider now the action
\bear{2M.1}
{2\kappa^{2}}
[ S[g,\phi,F^a] -  S_{\GHY} ] =
\int_{N} d^{D}z \sqrt{|g|} \{ {R}[g]
- C_{\alpha\beta}g^{MN} \partial_{M} \Phi^\alpha \partial_{N} \Phi^\beta
\nn\\
- \sum_{a \in \Delta}
\frac{\eta_a}{n_a!} \exp[ 2 \lambda_{a} (\Phi) ] (F^a)^2 \}
\ear
of a self-gravitating $\sigma$ model on $M$ with
topological term $S_{\GHY}$.
Here the $l$-dimensional target space, defined
by a vector field $\phi$ with scalar components
$\phi^\alpha$, $\alpha=1,\ldots,l$, is coupled to
several antisymmetric $n_a$-form fields $F^a$
via $1$-forms $\lambda_{a}$, $a\in\Delta$.
For consistency, we have to demand of course
that all fields are internally homogeneous.
We will see below how this gives rise
to an effective  $l+|\Delta|$-dimensional
target-space extension.
Note also that for convenience here we work with
fields $\phi$ and $F$ which differ from the
actual (physical) matter fields by a
rescaling with the square root of the coupling constant.

With $I\subset\{1,\ldots,n\}$, the generalized Maxwell fields
$F^a$ are located on $(n_a-1)$-dimensional world sheets
\bear{2M.20} M_{I} &:=& \prod_{i\in I} M_{i}= M_{i_1}  \times
\ldots \times M_{i_k}, \ear \bear{2M.19} n_a-1=D(I) &:=&  \sum_{i
\in I} d_i = d_{i_1} + \ldots + d_{i_k}. \ear of different
$(n_a-2)$-branes, labeled for each $a$ by the sets $I$ in a
certain subset $\Omega_a$ of the power set of $\{1,\ldots,n\}$.
Variation of (\ref{2M.1}) yields the field equations \bear{2M.4}
R_{MN} - \frac{1}{2} g_{MN} R  &=&   T_{MN}  ,
\\
\label{2M.5}
C_{\alpha\beta}{\Delta}[g] \phi^\beta -
\sum_{a \in \Delta} \frac{\eta_a \lambda^{\alpha}_a}{n_a!}
e^{2 \lambda_{a}(\phi)} (F^a)^2 &=& 0 ,
\\
\label{2M.6}
\nabla_{M_1}[g] (e^{2 \lambda_{a}(\phi)}
F^{a, M_1 M_2 \ldots M_{n_a}})  &=&  0 ,
\ear
$a \in \Delta$, $\alpha=1,\ldots,l$.

In (\ref{2M.4}) the $D$-dimensional energy-momentum
resulting from \rf{2M.1} is given by a sum
\bear{2M.7}
T_{MN} :=  \sum_{\alpha=1}^l T_{MN}[\phi^\alpha,g]
+ \eta_a \sum_{a\in\Delta} e^{2 \lambda_{a}(\phi)} T_{MN}[F^a,g] ,
\ear
of contributions from scalar and
generalized Maxwell fields,
\bear{2M.8}
T_{MN}[\phi^\alpha,g] &:=&
C_{\alpha\beta}\p_{M} \phi^\alpha \p_{N} \phi^\alpha -
\frac{1}{2} g_{MN} \p_{P} \phi^\alpha \p^{P} \phi^\alpha ,
\\
\label{2M.9}
T_{MN}[F^a,g] &:=&
\frac{1}{n_{a}!}  \left[
- \frac{1}{2} g_{MN} (F^{a})^{2}
 + n_{a}  F^{a}_{\ M M_2 \ldots M_{n_a}} F_{\ N}^{a\ M_2 \ldots M_{n_a}}
\right] .
\ear
We give now a sufficient criterion for the
energy-momentum tensor \rf{2M.7} to decompose
multidimensionally.

Let $W_1:= \{ i \mid i>0,\ d_i=1\}$ be the label set of
$1$-dimensional factor spaces of the multidimensional
decomposition, and set $n_1:=| W_1 |$. Define \beq{Wdef}
W(a;i,j):= \{ (I,J) \mid I,J\in \Omega_{a},\
                        (I\cap J) \cup \{i\} = I \not\ni j,\
                        (I\cap J) \cup \{j\} = J \not\ni i  \}
\eeq
Then the following holds.

{\bf Theorem:}
If for $n_1>1$ the $p$-branes satisfy
the condition
for all $a\in \Delta$, $i,j\in W_1$ with $i\neq j$,
the condition
\beq{2M.r2}
W(a;i,j) \mustbe \emptyset \quad \forall a\in \Delta \forall i,j\in W_1 ,
\eeq
then the energy-momentum \rf{2M.7}
decomposes multidimensionally without further constraints.

Proof: The only possible obstruction to the multidimensional
decomposition of  \rf{2M.7} comes from the  second term of
\rf{2M.9}, $F^{a}_{\ M M_2 \ldots M_{n_a}} F_{\ N}^{a\ M_2 \ldots
M_{n_a}}$ when the indices $M$ and $N$ take values in different
index sets labeling different $1$-dimensional factor spaces. The
theorem then follows just from the antisymmetry of the $F$-fields.
\hfill\mbox{$\Box$}\break

{\bf Corollary:}
A sufficient condition for
the multidimensional decomposition of  \rf{2M.7} is
\beq{2M.r1}
n_1 \mustbeleq 1 .
\eeq

If condition \rf{2M.r2} does not hold,
multidimensional decomposability of \rf{2M.7}
may impose additional non-trivial constraints
on the $p+2$-form fields.

Let us now specify the components of the $F$-fields
of generalized electric and magnetic type.

Antisymmetric fields of generalized electric type,
are given by scalar potential fields $\Phi^{a,I}$,
$a \in \Delta$, $I \in \Omega_{a}$,
which compose to a $(\sum_{a \in \Delta}|\Omega_{a}|)$-dimensional
vector field $\Phi$.
Magnetic type fields are just given as the duals of
appropriate  electric ones.
\bear{elmagF}
F^{e,I}&=& d\Phi^{e,I}\wedge\tau(I) \\
F^{m,I}&=& e^{-2\lambda_a(\phi)} * (d\Phi^{m,I}\wedge\tau(J)) .
\ear
In the Einstein frame, the action then reduces to
\bear{Sf0M}
&\! &\!{}^{(\E)}\!S[g^{(0)} ,\beta, \phi,\Phi]
= \int_{M_0} d^{D_0}x \sqrt{|g^{(0)}|}
\left\{\frac{m}{2}
\left[ R[g^{(0)}] - G_{ij}(\p\beta^i)(\p\beta^j)
\right.
\right.
\nn\\
&\! &\!
\ \quad
\left.
\left.
-C_{\alpha\beta} (\p \phi^\alpha) (\p \phi^\beta)
- \sum_{ {a\! \in\! \Delta},{I\! \in\! \Omega_{a}} }
\varepsilon_{a,I} e^{2 (\lambda_a(\phi) - d_i \beta^i )}
(\p \Phi^{a,I})^2
\right]
- {}^{(\E)}\!V(\beta)
\right\} \ , \ \quad
\ear
which corresponds to an purely Einsteinian $\sigma$-model
on  $M_0$ with extended
$(n + l + \sum_{a\in \Delta}| \Omega_{a}|)$-dimensional target space
and dilatonic potential (\ref{Vf0}).
Here and below we will consider by default the Einstein frame,
and set correspondingly $G_{ij}:={}^{(\E)}\!G_{ij}$.
In \rf{Sf0M} and below a summation convention is assumed
also on the extended target space.

For convenience, let us introduce the topological numbers
\beq{topnum}
l_{jI} := - \sum_{i \in I} D_i \delta^i_j ,
\qquad j =1,\ldots,n ,
\eeq
and with $N := n + l$ define
and define a $N \times |S|$-matrix
\beq{Lmat}
L = \left(L_{As} \right)
=
\left( \begin{array}{cc}
            &L_{i s} \\
             &L_{\alpha s}
             \end{array}
\right)
:=
\left( \begin{array}{cc}
            &l_{i I} \\
             &\lambda_{\alpha a}
             \end{array}
\right) ,
\eeq
a $N$-dimensional vector field
$(\sigma^A) := (\beta^i, \phi^{\alpha})$, $A=1,\ldots,n,n+1,\ldots,N$,
composed by dilatonic and matter scalar fields, and
a non-degenerate (block-diagonal) $N \times N$-matrix
\beq{Ghat}
\hat{G} = \left(\hat{G}_{AB} \right)    =
                             \left(
                              \begin{array}{cc}

                               G_{ij} &  0 \\
                                   0   &  C_{\alpha \beta}
                               \end{array}
                          \right) .
\eeq
With these definitions, \rf{Sf0M} takes the form
\bear{SGhat}
S_{0} =
\int_{M_0} d^{D_0}x \sq {g^{(0)} } \Bigl\{
\frac{m}{2}
\left[
{R}[g^{(0)}]
- \hat{G}_{AB} \p \sigma^A \p \sigma^B
- \sum_{s \in S} \varepsilon_s e^{2 L_{A s} \sigma^A}
(\p \Phi^s)^2
\right]
- {}^{(\E)}\!V(\sigma)
\Bigr\} \quad .
\ear

\section{\bf Solution with Abelian target-space}
\setcounter{equation}{0}
In the this section we consider the $\sigma$-model
\rf{Sf0M} without the $\Phi$ fields from the $p+2$-forms,
whence the target-space is the $n+l$-dimensional Abelian one,
and present a particularly interesting vacuum solution.

We derive
for $D_0\neq 2$ an new exact
Ricci flat multidimensional solution
for the effective $\sigma$-model \rf{Sf0M}
in the harmonic gauge $(2-D_0){\ol\gamma}\mustbe d_i\beta^i $
with zero potential \rf{Vf0}
and zero $\Phi$.
The field equation then read
\bear{fe1}
{G}_{ij} \p_{\mu} \beta^i \p_{\nu} \beta^j
+{C}_{\alpha\beta} \p_{\mu} \phi^\alpha \p_{\nu} \phi^\beta
= 0 ,
&\qquad&
\mu,\nu=0,\ldots, D_0-1 ,
\\
\label{fe2}
{}^{(\E)}\!{G}_{ij} {\Delta}[\ol g^{(0)}] \beta^j = 0 ,
&\qquad&
i=1,\ldots,n ,
\\
\label{fe3}
{C}_{\alpha\beta} {\Delta}[\ol g^{(0)}] \phi^\beta = 0 ,
&\qquad&
\alpha=1,\ldots,l .
\ear
In particular, we now solve these equations with flat
$({\ol M}_0,\ol g^{(0)})$.
In this case, there exist $g$-harmonic $\ol M_0$-coordinates $\tau^\mu$,
$\mu=0,\ldots, D_0-1$.
Let $g^{(0)}=e^{-2\ol\gamma}\eta_{\mu\nu}d\tau^\mu d\tau^\nu$.
In such harmonic coordinates,
equations \rf{fe2} and \rf{fe3} are solved by
\bear{linbeta}
\beta^i=b^i_\mu \tau^\mu + c^i ,
&\qquad&
i=1,\ldots,n ,
\\
\label{linphi}
\phi^\alpha=b^{n+\alpha}_\mu \tau^\mu + c^{n+\alpha} ,
&\qquad&
\alpha=1,\ldots,l .
\ear
We set
\beq{pvarphi}
\varphi^i_\mu:=\partltau{\mu} \varphi^i\ ,\quad \mu=0,\ldots, D_0-1\ .
\eeq
With \rf{linbeta},
the harmonic gauge condition reads
\bear{hc5}
A_\mu:={}^{(\E)}\!q \varphi^1_{\mu} = \sum_i d_i b^i_\mu \mustbe 0\ ,
&\qquad&
\mu=0,\ldots,D_0-1 .
\ear
With the harmonic gauge constraint \rf{hc5},
Eq. \rf{fe1}
then reads
\bear{constr}
\sum_{i=2}^{n} \varphi^i_{\mu} \varphi^i_{\nu}
+\sum_{\alpha\beta=1}^{l} {C}_{\alpha\beta}  b^\alpha_{\mu} b^\beta_{\nu}
&=&
\sum_{i=1}^{n} d_i b^i_{\mu} b^i_{\nu}
+\sum_{\alpha\beta=1}^{l} {C}_{\alpha\beta}  b^\alpha_{\mu} b^\beta_{\nu}
\mustbe 0 ,
\nn\\
&&\mu,\nu=0,\ldots,D_0-1\ .\
\ear
For convenience, one can set $c^A:=0$, $A=1,\ldots,n+l$.
Then $\gamma=0$, whence the harmonic coordinates are simultaneously
proper coordinates,  and the solution reads explicitly,
\beq{scalsol}
g=\eta_{\mu\nu} d\tau^\mu \otimes d\tau^\nu
+\sum_{i=1}^{n} e^{2b^i_\lambda\tau^\lambda} g^{(i)} ,
\eeq
with linear coefficients
$b^i_\mu$,
$i=1,\ldots, n$,
$\mu=0,\ldots, D_0-1$,
satisfying
$D_0$ linear constraints \rf{hc5}
(the harmonic gauge)
and ${D_0}^2$ quadratic constraints \rf{constr}
(the harmonic Wheeler-de Witt constraints).

This solution shows a generalized inflationary
behaviour, which extends the familiar notion of inflation
w.r.t. time, as in cosmology, to inflation w.r.t. the
internal degrees of freedom on the $D_0$-dimensional world manifold
of an extended object.
The constraint \rf{hc5} implies that the total $(D-D_0)$-
dimensional volume remains constant
(like in a steady state universe \cite{BG})
on the world manifold
$M_0$,
although here
(unlike the stationary case \cite{BG})
individual factor spaces
may undergo inflationary expansion or contraction
in particular directions on $M_0$.
In the standard cosmological case $D_0=1$, this solution agrees with
the one described in \cite{BlZ}.

\section{Orthobrane solutions with ${}^{(\E)}\!V=0$}
\setcounter{equation}{0}
Now we present a class of solutions with ${}^{(\E)}\!V=0$,
where the field equations read
\bear{feq1}
R_{\mu \nu}[g^{(0)}]  =
\hat{G}_{AB} \p_{\mu} \sigma^A \p_{\nu} \sigma^B
+ \sum_{s \in S} \eps_s e^{2 L_{As} \sigma^A}
\p_{\mu} \Phi^s  \p_{\nu} \Phi^s ,
&\qquad&
\mu,\nu=1,\ldots, D_0 ,
\\
\label{feq2}
\hat{G}_{AB} {\btu}[g^{(0)}] \sigma^B
-  \sum_{s \in S} \eps_s L_{As} e^{2 L_{Cs} \sigma^C} (\p \Phi^s)^2 = 0 ,
&\qquad&
A = 1,\ldots, N ,
\\ \label{feq3}
\p_{\mu} \left( \sqrt{|g^{(0)}|} g^{{(0)} \mu \nu} e^{2 L_{As}
\sigma^A} \p_{\nu} \Phi^s \right) = 0 ,
&\qquad&
s \in S .
\ear

For the Abelian part of the target space metric
we set $(\hat{G}^{AB}) := (\hat{G}_{AB})^{-1}$.
\beq{i3.21}
< X,Y > := X_A \hat{G}^{AB} X_{B} .
\eeq
For $s\in S$ let us now consider vectors
\beq{i3.20}
L_{s} = (L_{As}) \in \R^N .
\eeq
\nl
\noindent
{\bf Definition:}
A non-empty set $S$ is called an {\em orthobrane} index set,
iff there exists a family of real non-zero coefficients
$\{\nu_s\}_{s \in S}$, such that
\beq{i3.19}
< L_s,L_r > =
(L^{T} \hat{G}^{-1} L)_{sr} = - \eps_s (\nu_s)^{-2}\delta_{sr} ,
\qquad s,r \in S .
\eeq
For  $s \in S$  and  $A = 1, \ldots, N$, we set
\beq{i3.17}
\alpha^A_s := - \eps_s (\nu_s)^{2}  \hat{G}^{AB} L_{Bs} .
\eeq
{}\nl
Here, (\ref{i3.19}) is just an orthogonality condition for the
vectors $L_s$, $s \in S$.
Note that $< L_{s}, L_{s} >$ has just the opposite sign of
$\eps_s$, $s \in S$.
With the definition above,
we obtain an existence criterion for  solutions.
\nl
\noindent
{\bf Theorem:}
Let  $ S$ be an orthobrane index set
with coefficients \rf{i3.17}.
If for any  $s\in S$
there is a function  $H_s > 0$ on $M_0$ such that
\beq{i3.18}
{\btu}[g^{(0)}] H_s = 0 ,
\eeq
i.e. $H_s$ is harmonic on $M_0$,
then,
the field configuration
\bear{i3.13}
R_{\mu \nu}[g^{(0)}]  = 0 ,
&\qquad&
\mu,\nu=1,\ldots, D_0 ,
\\
\label{i3.14}
\sigma^A = \sum_{s \in S} \alpha^A_s \ln H_s ,
&\qquad&
A = 1, \ldots, N ,
\\
\label{i3.15}
\Phi^s  = \frac{\nu_s}{H_s} ,
&\qquad&
s \in S ,
\ear
satisfies the field equations (\ref{feq1})-(\ref{feq3}).
\hfill\mbox{$\Box$}\break
\nl
This theorem follows just from substitution of
(\ref{i3.19})-(\ref{i3.15}) into
the equations of motion (\ref{feq1})-(\ref{feq3}).
{}From (\ref{Ghat}), (\ref{Lmat}) and
(\ref{i3.21}) we get
\beq{i3.22}
< L_{s}, L_{r} > =  G^{ij} l_{iI} l_{jJ}
+ C^{\alpha\beta} {\lambda}_{\alpha a}  {\lambda}_{\beta b} ,
\eeq
with $s=(a,I)$ and $r=(b,J)$ in $S$
($a,b\in\Delta$, $I \in \Omega_{a}$,  $J \in \Omega_{b}$).
Here, the inverse of the dilatonic midisuperspace metric $G_{ij}$
is given by
\beq{3.26}
G^{ij} =
\frac{\delta_{ij}}{D_i} + \frac{1}{2 - D} ,
\eeq
whence, for $I, J \in \Omega$, with topological numbers $l_{iI}$
{}from \rf{topnum}, we obtain
\beq{i3.27}
 G^{ij} l_{iI} l_{jJ} = D(I \cap J) + \frac{D(I) D(J)}{2-D} ,
\eeq
which is again a purely topological number.

We set $\nu_{a,I} := \nu_{(a,I)}$. Then,
due to (\ref{i3.22}) and (\ref{i3.27}),
the orthobrane condition (\ref{i3.19}) reads
\beq{i3.28}
D(I \cap J) + \frac{D(I) D(J)}{2-D}
+ C^{\alpha\beta} {\lambda}_{\alpha a}  {\lambda}_{\beta b}
= - \eps(I) (\nu_{a,I})^{-2} \delta_{ab}\delta_{I,J} ,
\eeq
for $a, b \in \Delta$, $I \in \Omega_{a}$, $I \in \Omega_{b}$.
With $(a,I)=s\in S$, the coefficients (\ref{i3.17}) are
\bear{i3.29}
\alpha^i_s
&=& - \eps(I) G^{ij}l_{jI}  \nu_{a,I}^{2}
= \eps(I)
\biggl( \sum_{j \in I} \delta^i_j  + \frac{D(I)}{2-D} \biggr)
\nu_{a,I}^{2} ,
\qquad i = 1, \ldots, n ,
\\
\label{i3.30}
\alpha^\beta_s &=&
- \eps(I) C^{\beta\gamma}\lambda_{\gamma a}  \nu_{a,I}^{2} ,
\qquad \beta=1,\ldots,l
\ear
With $(\sigma^A) = (\phi^i, \varphi^\beta)$,
according to (\ref{i3.14}),
\bear{i3.31}
\beta^i = \sum_{s \in S} \alpha^i_s   \ln H_s ,
\qquad i = 1, \ldots, n ,
\\ \label{i3.32}
\phi^\beta = \sum_{s \in S} \alpha^\beta_s  \ln H_s ,
\qquad \beta=1,\ldots,l ,
\ear
and the harmonic gauge  reads
\beq{i3.33}
\gamma = \sum_{s \in S}
\alpha^0_s  \ln H_s ,
\qquad
\eeq
where
\beq{i3.34}
\alpha^0_s  := \varepsilon(I) \frac{D(I)}{2-D}  \nu_{a,I}^{2} .
\eeq
With $H_{a,I}:=H_{(a,I)}$, \rf{i3.29}, \rf{i3.30}, and \rf{i3.34},
the solution of \rf{i3.13} - \rf{i3.15} reads
\bear{xxxxx} \nn
g&=& \biggl(
\prod_{s \in S}
H_s^{2 \alpha^0_s}
\biggr)
g^{(0)}
+ \sum_{i=1}^{n}
\biggl(
\prod_{s \in S}
H_s^{2 \alpha^i_s}
\biggr)
g^{(i)}
\\
\label{i4.1}
&=&
\left(
\prod_{(a,I)\in S}
H_{a,I}^{ \eps(I) 2 D(I) \nu^2_{a,I} }
\right)^{1/(2-D)}
\left\{
g^{(0)}
+  \sum_{i=1}^{n}
\left(
\prod_{(a,I) \in S, I \ni i}
H_{a,I}^{\eps(I) 2 \nu^2_{a,I} }
\right)
g^{(i)}
\right\} ,
\\\nn 
&&
\mbox{\rm with}\
{\rm Ric}[g^{(0)}]=0,\qquad
{\rm Ric}[g^{(i)}] =0,\qquad i=1,\ldots n ,
\ear
\bear{i4.p}
\phi^\beta  =  \sum_{s \in S}
\alpha^\beta_{s} \ln H_s
= - \sum_{(a,I) \in S}
\eps(I) C^{\beta\gamma}\lambda_{\gamma a}
\nu^2_{a,I} \ln H_{a,I} ,
&\qquad& \beta=1,\ldots,l ,
\\
\label{i4.a1}
A^{a} = \sum_{I \in \Omega_{a}}
\frac{\nu_{a,I}}{H_{a,I}} \tau_{I} ,
&\qquad& a \in \Delta ,
\ear
where forms $\tau_I$ are defined in \rf{2.8},
parameters $\nu_s \neq 0$ and ${\lambda}_a$
satisfy the orthobrane condition (\ref{i3.28}),
$H_s$ are positive harmonic functions on $M_0$, and
${\rm Ric}[g^{(i)}]$ denotes the Ricci-tensor of $g^{(i)}$.
Finally recall that these solutions
are subject to the  {\em orthobrane} constraints
\beq{i4.2}
D(I \cap J) + \frac{D(I) D(J)}{2-D}
+ C^{\alpha\beta} {\lambda}_{\alpha a}  {\lambda}_{\beta b}
= - \eps(I) (\nu_{a,I})^{-2} \delta_{ab}\delta_{I,J}\ ,
\quad 0\neq\nu_{a,I}\in\R\ ,
\eeq
for $a, b \in \Delta$, $I \in \Omega_{a}$, $I \in \Omega_{b}$.
These condition lead to specific intersection rules
for the $p$-branes involved.
Some concrete examples
of {\em orthobrane} solutions have  been elaborated
in \cite{IMR}.

For positive definite $(C_{\alpha\beta})$
(or $(C^{\alpha \beta})$) and $D_0 \geq 2$, \rf{i4.2} implies
\beq{i3.43}
\eps(I)=-1 ,
\eeq
for all $I \in \Omega_{a}$, $a \in \Delta$.
Then, the restriction $g_{\vert M_{I}}$ of the metric \rf{i4.1}
to a membrane manifold $M_{I}$ has an odd number
of negative eigenvalues,
i.e. linearly independent time-like directions.
However, if the metric $(C_{\alpha\beta})$ in the space of scalar fields
is not positive definite,
then \rf{i3.43} may be violated for sufficiently negative
$C^{\alpha\beta} {\lambda}_{\alpha a}  {\lambda}_{\beta b}<0$.
\section{\bf Target space structure}
\setcounter{equation}{0}

{\bf Theorem:}
The target space $({\frak M},{\frak g})$ is a homogeneous space.

Proof: The Killing vectors of a transitive subgroup
of $\Isom({\frak M})$
can be determined explicitly.
\bear{KVtransit}
V_s &:=& \frac{\p}{\p \Phi^s},
\qquad
s\in S,
\nn\\
U_A &:=&
\frac{\p}{\p x^A}
- \sum_{s \in S} U^{s}_ A \Phi^s \frac{\p}{\p \Phi^s},
\qquad
A=1,\ldots,N.
\ear

Moreover, the Lie-algebra of
the transitive group of isometries
generated by \rf{KVtransit} reads
\bear{Liealg}
[U,U] = [V,V] &= & 0
\nn\\ \label{noncom}
[U_A, V_s] &=&  L^s_A V_s ,
\qquad A=1,\ldots, N, \quad  s\in S\ .
\ear

{\bf Theorem:}
The target space $({\frak M},{\frak g})$ is locally symmetric
if and only if $<L^s,L^r>_{\hat G}(L^s-L^r)=0$.

Proof: Let ${\frak Riem}$ denote the Riemann tensor of
$({\frak M},{\frak g})$.
The latter is locally symmetric, if and only if
\beq{locsym}
\nabla{\frak Riem}=0 ,
\eeq
where $\nabla$
denotes the covariant derivative w.r.t ${\frak g}$.
However, the only non-trivial equations \rf{locsym} are
\beq{DRiem}
\nabla_p {\frak R}_{s r q A} =
k_{psrq}
<L^{s}, L^{r}>_{\hat G}(L^{r}_A - L^{s}_A)
= 0,
\quad
A=1,\ldots,N,
\quad
p,q,r,s\in S
\eeq
with
$k_{psrq}:=
\eps_{s} \eps_{r}
e^{2U^{s} + 2U^{r}}
(\delta_{p s} \delta_{r q} + \delta_{p r} \delta_{s q})$
nonzero for fixed $s,r$.

\section{\bf Scalar plus $p$-branes with spherical symmetry}
\setcounter{equation}{0}

let us now examine static, spherically symmetric, multidimensional
space-times with
\beq{8}
        M=M_{-1} \times M_0\times M_1\times \cdots \times M_N,
        \cm \dim M_i=d_i, \quad i=0,\ldots, N ,
\eeq
where $M_{-1}\subset {\R}$ corresponds to a radial coordinate $u$,
$M_0 = S^2$ is a 2-sphere, $M_1\subset\R$ is time, and $M_i,\ i>1$
are internal factor spaces.
The metric is assumed correspondingly to be
\bear{9}
     ds^2 \eql
            \e^{2\alpha(u)}du^2 + \sum_{i=0}^N\e^{2\beta_i(u)}ds_i^2
\nn\\
            \al\equiv\al
                -\e^{2\gamma(u)}dt^2+\e^{2\alpha(u)}du^2
              +\e^{2\beta_0(u)}d\Omega^2
              +\sum_{i=2}^N\e^{2\beta_i(u)}ds_i^2  ,
\ear
where
$ds_0^2 \equiv d\Omega^2=d\theta+\sin^2\theta\, d\phi^2$
is the line element on $S^2$,
$ds_1^2 \equiv -dt^2$ with $\beta_1 =: \gamma$,
and $ds_i^2$, $i> 1$, are $u$-independent line elements of internal
Ricci-flat spaces of arbitrary dimensions $d_i$ and signatures
$\eps_i$.

For simplicity here let us only consider a single
scalar field denoted as $\varphi$.

An electric-type $p+2$-form $F_{\e I}$ has a domain
given by a product manifold
\beq{MI}
        M_I = M_{i_1} \times \cdots \times M_{i_k} ,
\eeq
where
\beq{Ie}
        I = \{i_1,\ldots,i_k\} \subset I_0 \eqdef \{0,1,\ldots,N\} .
\eeq
The corresponding dimensions are
\beq{d(I)}
        d(I) \eqdef \sum_{i\in I} d_i,\inch    d(I_0) = D-1.
\eeq

A magnetic-type $F$-form of arbitrary rank $k$ may be defined
as a form on a  domain $M_{\oI}$ with $\oI \eqdef I_0 - I$,
dual to an electric-type form,
\beq{Fm}
     F_{\m I,\,M_1\ldots M_k}
        = \e^{-2\lambda\varphi} (*F)_{\e I,\,M_1\ldots M_k}
        \equiv \e^{-2\lambda\varphi}
        \frac{\sqrt{g}}{k!} \eps_{M_1\ldots M_k N_1\ldots N_{D-k}}
        F_{\e I}^{N_1\ldots N_{D-k}},
\eeq
where $*$ is the Hodge operator and
$\eps$ is the totally antisymmetric Levi-Civita symbol.

For simplicity we now considering a just a single $n$-form,
i.e. a single electric type and a single dual magnetic component,
whence
\beq{11}
        \rank F_{\m I} = D - \rank F_{\e I} = d(\oI),
\eeq
whence $k=n$ in (\ref{Fm}) and
\beq{12}
        d(I) = n-1   \quad \mbox{for} \quad F_{\e I}, \cm
        d(I) = d(I_0)-n = D-n-1 \quad \mbox{for} \quad F_{\m I}.
\eeq

All fields must be compatible
with spherical symmetry and staticity.
Correspondingly, the vector $\varphi$ of scalars and
the $p+2$-forms valued fields depend (besides on their domain as forms)
on the radial variable $u$ only.

Furthermore,
the domain of the electric form $F_{\e I}$
does not include the sphere $M_0 = S^2$, and
$F_{\e I}$
is specified by a
$u$-dependent potential form,
\beq{Ue}
        F_{\e I,\,u L_2\ldots L_n} = \bdy_{\,[u}U_{L_2\ldots L_n ]} \cm
        U = U_{L_2,\ldots,L_{n}} dx^{L_2}\wedge\ldots\wedge dx^{L_n} .
\eeq

Since the time manifold $M_1$ is a factor space of $M_I$, the form
(\ref{Ue}) describes an electric $(n-2)$-brane in the remaining
subspace of $M_I$. Similarly \rf{Fm} describes a magnetic
$(D-n-2)$-brane in $M_I$.

Let us label all nontrivial components of $F$ by a collective
index $s = (I_s,\chi_s)$, where $I=I_s\subset I_0$ characterizes the
subspace of $M$ as described above and
$\chi_s=\pm 1$ according to the rule
\beq{13}
        \e \mapsto \chi_s =+1, \inch \m \mapsto \chi_s = -1.
\eeq
If $1\in I$, the corresponding $p$-brane evolves with $t$
and we have a true electric or magnetic field,
otherwise
the potential \rf{Ue} does not depend on $\ol M_0$,
i.e. it is just a scalar in 4 dimensions.
In this case we call the corresponding electric-type $F$ component
\rf{Ue} {\em electric quasiscalar} and its dual,
magnetic-type, $F$ component \rf{Fm} {\em magnetic quasiscalar}.
So there are in general four types of $F$-field components
(summarized in Table \ref{tab1}): electric (\E), magnetic (\M),
electric quasiscalar (\EQ), magnetic quasiscalar (\MQ).
\medskip
\begin{table}[htb!]
\caption{Different types of antisymmetric $p+2$-form  fields}
\label{tab1}
\begin{center}
\begin{tabular}{|l|lll|}
\hline
 & &  &
\\
\E   &  electric ($1\in I$) & $F_{tuA_3\ldots A_n}$  &
       $A_k$ (coordinate) index of $M_I$
\\
\M   &  magnetic ($1\in I$)  & $F_{\theta\phi B_3\ldots B_n}$ &
       $B_l$ (coordinate) index of $M_{\oI}$
\\
 & & &
\\
\EQ  &  electric quasiscalar ($1\not\in I$) &  $F_{uA_2\ldots A_n}$ &
      $A_k$ (coordinate) index of  $M_{I}$
\\
\MQ  &  magnetic quasiscalar ($1\not\in I$) & $F_{t\theta\phi B_4\ldots B_n}$ &
      $B_l$ (coordinate) index of  $M_{\oI}$
\\
 &  &  &
\\
\hline
\end{tabular}
\end{center}
\end{table}
The choice of subsets $I_s$ is only constrained by
the multidimensional decomposition condition \ref{2M.r2} for
the energy-momentum tensor.
Since antisymmetric $p+2$-form  field components of
type \E\ and \M\ (and type \EQ\ and \MQ\ respectively)
just complement each other, they
should be considered as independent of each other.
In the following we
consider all $F_s$ as independent fields
(up to index permutations) each with
a single nonzero component.

Let us assume Ricci-flat internal spaces.
With spherical symmetry and staticity
all field become independent of $M_{0}$ and $M_{0}$
respectively.
And the variation reduces further from $\ol M_0$ to the
radial manifold $M_{-1}$.

The reparametrization gauge on the lower dimensional manifold
here is chosen as the (generalized) harmonic one \cite{Ra0}.
Since $M_{-1}$ is $1$-dimensional $u$ is a harmonic coordinate,
$\DAL u = 0$, such that
\beq{3.1}
        \alpha (u)=\sigma_0 (u).
\eeq
The nonzero Ricci tensor components are
\bear{Ricci}
        \e^{2\alpha}R_t^t   \eql  -\gamma'',
\nn\\
        \e^{2\alpha}R_u^u   \eql -\alpha''+{\alpha'}^2-{\gamma'}^2
                       -2{\beta'}^2-\sum_{i=2}^N d_i{\beta'_i}^2,
\nn\\
     \e^{2\alpha}R_\theta^\theta   \eql
                \e^{2\alpha}R_\phi^\phi= \e^{2\alpha-2\beta}-\beta'',           \nn
\nn\\
     \e^{2\alpha}R_{a_j}^{b_i} \eql
               -\delta_{a_j}^{b_i}\beta''_i \inch   (i,j=1,\ldots,N)\,,
\ear
where a prime denotes $d/du$ and the indices $a_i,\ b_i$ belong to the
$i$-th internal factor space. The Einstein tensor component $G_1^1$ does
not contain second-order derivatives:
\beq{G11}
     \e^{2\alpha}G_1^1=-\e^{2\alpha-2\beta}
    +\Half {\alpha'}^2-\Half \biggl({\gamma'}^2
    +2{\beta'}^2+\sum_{i=2}^N d_i{\beta'_i}^2\biggr).
\eeq
The corresponding component of the Einstein equations is an integral of
other components, similar to the energy integral in cosmology.

The generalized Maxwell equations 
give
\bear{3.2}
        F_{\e I}^{uM_2\ldots M_n}
                \eql Q_{\e I}\e^{-2\alpha - 2\lambda\varphi},
                        \qquad \ \ Q_{\e I}= \const,
\\\label{3.3}
        F_{\m I,\, uM_1\ldots M_{d(\oI)}}
                \eql Q_{\m I} \sqrt{|g_{\oI}|},\qquad
                        \qquad  Q_{\m I}= \const,
\ear
where $|g_{\oI}|$ is the determinant of the $u$-independent
part of the metric of $M_{\oI}$ and $Q_s$ are charges.
These solutions provide then the energy momentum tensors,
of the electric and magnetic $p+2$-forms written in matrix form,
\bearr{3.4x}
    \e^{2\alpha}(T_M^N [F_{\e I}])
          = -\half \eta_F \eps(I) \Qie^2 \e^{2y_{\e I}}
                 \diag\bigl(+1,\ \pO_I,\ \mO_{\oI} \bigr);
\nnnv
    \e^{2\alpha}(T_M^N [F_{\m I}])
          = \half \eta_F \eps(\oI) \Qim^2
              \e^{2y_{\m I}} \diag\bigl(1,\ \pO_I,\ \mO_{\oI} \bigr),
\ear
where the first position belongs to $u$
and $f$ operating over $M_J$ is denoted by
$[f]\mathstrut_J$.
The functions $y_s (u)$ are
\beq{3.5}
        y_s (u) = \sigma (I_s) - \chi_s \lambda\varphi.
\eeq
The scalar field EMT 
is
\beq{3.6}
        \e^{2\alpha}T_M^N [\varphi]
            = \half {(\varphi^a)'}^2 \diag\bigl(+1,\ \mO_{I_0}\bigr).
\eeq

The sets $I_s\in I_0$ may be classified by types \E, \M, \EQ, \MQ\ according
to the description in the previous section. Denoting $I_s$ for the
respective types by $I_\E,\ I_\M,\ I_{\EQ},\ I_{\MQ}$,
we see from (\ref{3.4x})
that,  positive electric and
magnetic energy densities require
\beq{3.7}
     \eta_f = -\eps(I_\E) = \eps(\oI_\M) = \eps(I_{\EQ}) = -\eps(\oI_{\MQ}).
\eeq
If $t$ is the only time coordinate, (\ref{3.7}) with $\eta_F=1$
holds for any choices of $I_s$. If there exist other times,
then the relations (\ref{3.7})
constrain the
subspaces where the different $F$ components may be specified.

Since the total EMT on the r.h.s. of the Einstein equations  has
the property
\beq{3.8}
        T_u^u + T_\theta^\theta = 0 ,
\eeq
the corresponding combination on the l.h.s
becomes an integrable Liouville form
\bear{3.9}
        G_u^u + G_\theta^\theta \eql \e^{-2\alpha}
            \bigl[-\alpha'' + \beta''_0
               + \e^{2\alpha - 2\beta_0} \bigr] = 0,
\nnv
        \e^{\beta_0 - \alpha} \eql s(k,u),
\ear
where $k$ is an integration constant (IC) and the function $s(k,.)$
is defined as follows:
\bear{3.10}
        s(k,u) \eqdef \vars{ k^{-1} \sinh ku, \quad & k>0
\\
                                           u,       & k=0
\\
                                     k^{-1} \sin ku,     & k<0
                            }
\ear
Another IC is suppressed by adjusting the origin of the $u$
coordinate.

With (\ref{3.9}) the $D$-dimensional line element may be written in the
form
\beq{3.11}
        ds^2 = \frac{\e^{-2\sigma_1}}{s^2(k,u)}
                  \biggl[ \frac{du^2}{s^2(k,u)} + d\Omega^2\biggr]
                                + \sum_{i=1}^{N} \e^{2\beta_i}ds_i^2
\eeq
where $\sigma_1$ has been defined in (\ref{e14}).

Let us treat the whole set of unknowns ${\beta_i(u),\ \varphi(u)}$
as a real-valued vector function $x (u)$ in an $(N+1)$-dimensional
vector space $V$, with components $x^A = \beta_A$ for $A=1,\ldots,
N$ and $x^{N+1}=\varphi$.

Then the field equations for
$\beta_i$ and $\varphi$ coincide with the equations of motion
corresponding to the Lagrangian of a Euclidean Toda-like system
\bear{3.13}
        L=\oG_{AB}{x'}^A {x'}^B -  V_Q (y) , \inch
                V_Q (y) = \sum_s \theta_s Q_s^2 \e^{2y_s} ,
\ear
where, according to (\ref{3.7}),
$\theta_s=1$ if $F_s$ is a true electric or magnetic field
and $\theta_s =-1$ if $F_s$ is quasiscalar.
The nondegenerate, symmetric matrix
\beq{3.14}
(\oG _{AB})=\pmatrix {
        G_{ij}&  0      \cr
        0     &  1      \cr }, \inch
                        G_{ij} = d_id_j + d_i \delta_{ij}
\eeq
defines a positive-definite metric in $V$.
The energy constraint corresponding to \rf{3.13} is
\beq{3.15}
        E = {\sigma'_1}^2 + \sum_{i=1}^{N} d_i {\beta'}_i^2
                         +\varphi'^2 + V_Q (y)
          = \oG_{AB}{x'}^A {x'}^B + V_Q (y)= 2k^2\sign k,
\eeq
with $k$ from (\ref{3.9}).
The integral (\ref{3.15}) follows here from the $(uu)$-
component of (\ref{2M.4}).

The functions $y_s(u)$ (\ref{3.5}) can be represented as scalar
products in $V$ (recall that $s = (I_s, \chi_s)$):
\bear{3.16}
        y_s (u)   = Y_{s,A}  x^A,    \inch
        (Y_{s,A}) = (d_i\delta_{iI_s}, \ \  -\chi_s \lambda),
\ear
where $\delta_{iI} := \sum_{j\in I}\delta_{ij}$
is an indicator for $i$ belonging to $I$ (1 if $i\in I$ and 0 otherwise).

The contravariant components of $Y_s$ are found using the matrix
$\oG^{AB}$ inverse to $\oG_{AB}$:
\bearr{3.18}
        (\oG ^{AB})=\pmatrix{
                G^{ij}&      0      \cr
                0     &      1      \cr }, \inch
        G^{ij}=\frac{\delta^{ij}}{d_i}-\frac1{D-2} \\ \lal
                                                                                                 \label{3.19}
        (Y_s{}^A) =
        \biggl(\delta_{iI_s}-\frac{d(I_s)}{D-2}, \ \ -\chi_s \lambda\biggr),
\ear
and the scalar products of different $Y_s$, whose values are of primary
importance for the integrability of our system, are
\beq{3.20}
        Y_{s,A}Y_{s'}{}^A = d(I_s \cap I_{s'})
                                        - \frac{d(I_s)d(I_{s'})}{D-2}
                        + \chi_s\chi_{s'} \lambda^2.
\eeq

\section{Purely EM black hole solutions}   
\setcounter{equation}{0}
In \cite{BKR} it was shown that quasiscalar components of the
$F$-fields are incompatible with orthobrane \bhs. Therefore let us
now consider only two $F$-field components, Type \E\ and Type \M\
according to the classification above. They will be electric as
$F_\e$ and $F_\m$ and the corresponding sets $I_s\subset I_0$ as
$I_\e$ and $I_\m$. Then a minimal configuration (\ref{8}) of the
manifold $M$ compatible with an arbitrary choice of $I_s$ has the
following form:

\beq{6.1}
        N=5,\cm I_0 = \{0,1,2,3,4,5\},  \cm
        I_e = \{1,2,3\},        \cm
        I_m = \{1,2,4\},
\eeq
so that
\bearr
        d(I_0) = D-1, \qquad d(I_\e) = n-1, \qquad d(I_\m)= D-n-1,  \qquad
        d(I_\e \cap I_\m) = 1 + d_2;
\nnnv
\label{6.2}
        d_1=1, \inch   d_2+d_3 = d_3 + d_5 = n-2.
\ear
The relations (\ref{6.2}) show that, given $D$ and $d_2$, all $d_i$ are
known.

This corresponds to an electric
($n-2$)-brane located on the subspace $M_2\times M_3$ and a
magnetic ($D-n-2$)-brane on the subspace $M_2\times M_4$.
Their intersection dimension $d_{\ints}=d_2$ turns out to
determine qualitative  properties of the solutions.

The index $s$ now takes the two values $\e$ and $\m$ and
\bear{6.2a}
        Y_{\e, A} \eql (1, d_2, d_3, 0, 0, -\lambda);
\nn\\
        Y_{\m, A} = (1, d_2, 0, d_4, 0, \lambda);
\nn\\
     Y_\e^A     \eql (1,1,1,0,0, -\lambda)
                           - \frac{n-1}{D-2} (1,1,1,1,1,0);
\nn\\
        Y_\m^A    \eql (1,1,0,1,0, \lambda)
                                 - \frac{D-n-1}{D-2} (1,1,1,1,1,0),
\ear
where the last component of each vector refers to
$x^{N+1} = x^6 = \varphi$.

In the solutions presented below the set of
ICs will be reduced by the condition that the
space-time be asymptotically flat at spatial infinity $(u=0)$ and by
a choice of scales in the relevant directions. Namely, we put
\beq{6.3}
        \beta_i(0)=0=\varphi(0)    \cm i=1,2,3,4,5.
\eeq
The requirement $\varphi(0)=0$ is convenient and may be always
satisfied by a redefinition of the charges. The conditions $\beta_i(0)=0$
($i>1$) mean that the real scales of the extra dimensions are hidden in
the internal metrics $ds_i^2$ independent of whether or not they are
assumed to be compact.

In the following, both cases, orthobrane solutions and
solutions with degenerate charges, are considered first generally and
then for the minimal configuration \rf{6.1}-\rf{6.3}.

\subsection{Orthobrane black hole solutions}

Assuming that the vectors $Y_s$ are mutually orthogonal with
        respect to the metric $\oG_{AB}$, i.e.
\beq{4.1}
        Y_{s,A}Y_{s'}{}^A = \delta_{ss'}N_s^2 ,
\eeq
the number of functions $y_s$ does not
exceed the number of equations,
and the system becomes integrable.
Due to (\ref{12}), the
norms $N_s$ are actually $s$-independent:
\beq{4.1a}
        N_s^2 = d(I_s)\biggl[1- \frac{d(I_s)}{D-2} \biggr] + \lambda^2
        = \frac{(n-1)(D-n-1)}{D-2} + \lambda^2 \eqdef \frac{1}{\nu},
\eeq
$\nu >0$.

Due to (\ref{4.1}), the functions $y_s(u)$ obey the decoupled equations
\beq{4.2}
        y''_s = \theta_s \frac{Q_s^2}{\nu} \e^{2y_s},
\eeq
whence
\beq{4.3}
        \e^{-y_s(u)} = \vars{
                (|Q_s|/\sqrt{\nu}) s(h_s,\ u+u_s),  & \theta=+1,   \yy
        [|Q_s|/(\sqrt{\nu} h_s)] \cosh[h_s(u+u_s)],\qquad h_s>0, \quad
                                                             & \theta=-1.
                                  }
\eeq
where $h_s$ and $u_s$ are ICs and the function $s$ was defined in
(\ref{3.10}). For the functions $x^A (u)$ we obtain:
\beq{4.4}
        x^A(u) = \nu \sum_s Y_s{}^A y_s(u) + c^A u + \oc^A,
\eeq
where the vectors of ICs $c^A$ and $\oc^A$ satisfy the orthogonality
relations $c^A Y_{s,A} = \oc^A Y_{s,A} = 0$, or
\beq{4.5}
        c^i   d_i\delta_{iI_s} - \lambda  c^{N+1}\chi_s =0, \inch
        \oc^i d_i\delta_{iI_s} - \lambda \oc^{N+1}\chi_s=0.
\eeq
Specifically, the logarithms of the scale factors $\beta_i$ and the
scalar field $\varphi$ are
\bear{4.6}
        \beta_i (u) \eql
                         \nu \sum_s \biggl[
     \delta_{iI_s} - \frac{d(I_s)}{D-2}\biggr] y_s (u)
                        +c^i u + \oc^i,
\\   \label{4.7}
        \varphi (u)\eql
                    - \lambda\nu \sum_s y_s(u) + c^{N+1} u + \oc^{N+1},
\ear
and the function $\sigma_1$ which appears in the metric (\ref{3.11}) is
\bearr{4.8}
        \sigma_1 = -\frac{\nu}{D-2}\sum_s d(I_s)\, y_s(u) + c^0 u + \oc^0
\ear
with
\beq{4.8a}
        c^0 =  \sum_{i=1}^{N} d_i c^i,  \inch
        \oc^0 =  \sum_{i=1}^{N} d_i \oc^i.
\eeq

Finally, (\ref{3.15}) now reads
\beq{4.9}
        E = \nu \sum_s h_s^2\sign h_s + \oG_{AB}c^A c^B
                                                   = 2k^2 \sign k.
\eeq

The relations (\ref{3.1}), (\ref{3.2}), (\ref{3.3}), (\ref{3.9}),
(\ref{3.11}), (\ref{4.3})--(\ref{4.9}), along with the definitions
(\ref{3.10}) and (\ref{4.1a}) and the restriction (\ref{4.1}), entirely
determine the general solution.

For the minimal configuration  \rf{6.1}-\rf{6.3},
the orthogonality condition (\ref{4.1})  reads
\beq{6.4}
        \lambda^2=  d_2+1 -\frac{1}{D-2}(n-1)(D-n-1)
\eeq In particular, in dilaton gravity $n=2,\ d_2=0$ and the
integrability condition \rf{6.4} just reads $\lambda^2 = 1/(D-2)$,
which is a well-known relation of string gravity. The familiar
Reissner-Nordstr\"om solution, $D=4,\ n=2$, $\lambda=0$, $d_2=0$
does {\em not} satisfy \eq(\ref{6.4}). (It will be recovered
indeed as a degenerate case below.) Some examples of
configurations satisfying the orthogonality condition (\ref{6.4})
in the purely topological case $\lambda=0$ are summarized in Table
\ref{tab2} (including the values of the constants $B$ and $C$ from
(\ref{6.12}) ). In this case (\ref{6.4}) is just a Diophantus
equation for $D$, $n$ and $d_2$.

\begin{table}[htb!]
\caption{Orthobrane solutions with $\lambda=0$}
\label{tab2}
\begin{center}
\begin{tabular}{|l|l|l|l|l|l|l|}
\hline
                &&&&&& \\
     & $n$   &  $d(I_{\e})$  &  $d(I_{\m})$  & $d_2$ & $B$ & $C$ \\
                        &&&&&& \\
      \hline
      &&&&&& \\
 $D = 4m + 2$        & 2m{+}1  & 2m &  2m &\ \ m{-}1 & 1/m  & 1/m\\
 ($m\in \N$) &&&&&& \\
                        &&&&&&                             \\
 \qquad D= 11          &  4    &  3 &   6  &    1 & 2/3  &  1/3   \\
                             &  7    &  6 &   3  &    1 & 1/3  &  2/3   \\
                        &&&&&&                             \\
\hline
\end{tabular}
\end{center}
\end{table}

The solution is entirely determined by
inserting (\ref{6.2a}) into
(\ref{4.4}) with $\oc^A=0$ due to (\ref{6.3}),
\beq{6.5}
        x^A(u) = \nu \sum_s Y_s{}^A y_s(u) + c^A u;  \cm
        \e^{-y_s(u)} = (|Q_s|/\sqrt{\nu}) s(h_s,\ u+u_s).
\eeq
Due to (\ref{6.4}) the parameter $\nu$ is
\beq{6.5a}
        \nu = 1/\sqrt{1+d_2}.
\eeq

The constants are connected by the relations
\bearr{6.6}
        \bigl(|Q_{\e,\m}|/\nu\bigr)\, s(h_{\e,\m}, u_{\e,\m}) =1;
\nnnv
        c^1 + d_2 c^2 + d_3 c^3 - \lambda c^6 =0;  \inch
        c^1 + d_2 c^2 + d_4 c^4 + \lambda c^6 =0;
\nnnv
        \frac{h_{\e}^2\sign h_{\e} + h_{\m}^2 \sign h_{\m}}{1+d_2}
          + G_{ij} c^i c^j + (c^6)^2  = 2k^2 \sign k,
\ear
where the matrix $G_{ij}$ is given in (\ref{3.14}) and
all $\oc^A =0$ due to the boundary conditions (\ref{6.3}).
The fields $\varphi$ and $F$ are given by \eqs (\ref{3.2}),
(\ref{3.3}), (\ref{4.7}).

This solution contains 8 nontrivial,
independent ICs, namely, $Q_\e,Q_\m,h_\e,h_\m$
and 4 others from the set $c^A$ constrained by \rf{6.6}.

For black holes,
we require that all $|\beta_i| < \infty$, $i= 2,\ldots,N$
(regularity of extra dimensions), $|\varphi| < \infty$ (regularity of
the scalar field) and $|\beta_0| < \infty$ (finiteness of the spherical
radius) as $u \to \infty$.
With $y_s(u) \sim -h_s u$, this leads to the following constraints on
the ICs:
\beq{5.3}
        c^A = -k \sum_s \Bigl(\delta_{1I_s} + \nu Y_s{}^A h_s\Bigr),
\eeq
where $A=1$ corresponds to $i=1$.
Via orthonormality
relations (\ref{4.5})  for $c^A$, we obtain
\bear{5.4}
        h_s \eql k \delta_{1I_s},
\yy\label{5.5}
        c^A \eql -k \delta^A_1
        +  k \nu \sum_s \delta_{1I_s} Y_s{}^A ,
\ear
and (\ref{4.9}) then holds
automatically.

Let us now consider the case where
(\ref{5.4}) and (\ref{5.5}) with $\delta_{1I_s} =1$ hold.
After a transformation
$u\mapsto R$, to isotropic coordinates given by the relation
\beq {6.7}
     \e^{-2ku}=1-2k/R ,
\eeq
we obtain
\bearr{6.8}
     ds^2 = -\frac{1-2k/R}{P_\e^{B}P_\m^{C}} dt^2
                +P_\e^{C}P_m^{B}
     \left(\frac{dR^2}{1-2k/R}+R^2 d\Omega^2\right)
                                +\sum_{i=2}^5 \e^{2\beta_i(u)}ds_i^2 ,
\yyy\label{6.9}
     \e^{2\beta_2} = {P_\e}^{-B}{P_\m}^{-C},
\cm
     \e^{2\beta_3} = \left({P_\m}/{P_\e}\right)^{B},
\nnnv
     \e^{2\beta_4} = \left({P_\e}/{P_\m}\right)^{C} ,
\cm
     \e^{2\beta_5} = {P_\e}^{C}{P_\m}^{B} ,
\yyy
\label{6.10}
 e^{2\lambda\varphi} =
     ({P_\e}/{P_\m})^{2\lambda^2/(1+d_2)} ,
 \yyy \label{6.11}
     F_{01M_3\ldots M_n}=-{Q_\e}/{(R^2 P_\e)} ,
\cm
     F_{23M_3 \ldots M_n}=Q_\m \sin\theta,
\ear
with the notations
\bearr{6.12}
        P_{\e,\m} = 1+ p_{\e,\m}/R, \qquad
        p_{\e,\m} = \sqrt{k^2+ (1+d_2) Q_{\e,\m}^2} -k;
\nnnv
        B = \frac{2(D-n-1)}{(D-2)(1+d_2)},
\cm
        C = \frac{2(n-1)}{(D-2)(1+d_2)}.
\ear

The BH gravitational mass as determined from a comparison of \rf{6.8}
with the Schwarzschild metric for $R\to \infty$ is
\beq{6.13}
     G_N M= k + \Half (B p_\e + C p_\m) ,
\eeq
where $G_N$ is the Newtonian gravitational constant.
This expression, due to $k>0$, provides a restriction
upon the charge combination for a given mass, namely,
\beq{6.14}
     B |Q_\e| +  C |Q_\m| < 2G_N M/\sqrt{1+d_2} .
\eeq
The inequality is replaced by equality in the extreme limit $k=0$.
For $k=0$ our BH turns into a naked singularity (at the centre $R=0$)
for any $d_2>0$, while for $d_2=0$ the zero value of $R$ is not a
centre ($g_{22}\neq 0$) but a horizon. In the latter case, if $|Q_e|$
and $|Q_m|$ are different, the remaining extra-dimensional scale
factors are smooth functions for all $R\geq 0$.

For a  static, spherical BH
one can define a Hawking temperature $T_H:= \kappa/2\pi$
as given by the surface gravity $\kappa$.
With a generalized Komar integral (see e.g. \cite{Heusler})
\beq{Komar}
M(r):=-\frac{1}{8\pi}
\int_{S_{r}} * d\xi
\eeq
over the time-like Killing form $\xi$,
the surface gravity can be  evaluated as
\beq{6.15}
     \kappa   = M(r_H)/(r_H)^2
     =(\sqrt{|g_{00}|})'\Big/\sqrt{g_{11}}\biggr|_{r=r_H}
     =\e^{\gamma-\alpha}|\gamma'|\, \biggr|_{r=r_H}\,,
\eeq
where a prime, $\alpha$, and $\gamma$ are understood in the sense of the
general metric \rf{9} and $k_{\rm B}$ is the Boltzmann constant.
The expression \rf{6.15} is invariant with respect to radial coordinate
reparametrization, as is necessary for any quantity having a direct
physical meaning.
It is also
invariant under conformal mappings
with a conformal factor which is smooth at
the horizon.

Substituting $g_{00}$ and $g_{11}$ from \rf{6.8}, one obtains:
\beq{6.16}
     T_H = \frac{1}{2\pi k_{\rm B}} \frac{1}{4k}
         \left[\frac{4k^2}{(2k+p_\e)(2k+p_\m)}\right]^{1/(d_2+1)}.
\eeq
If $d_2=0$ and both charges are nonzero, this temperature tends to zero
in the extreme limit $k\to 0$;  if $d_2=1$ and both
charges are nonzero, it tends to a finite limit, and in all other cases
it tends to infinity.  Remarkably, it is determined by the
$p$-brane intersection dimension $d_2$ rather than the whole space-time
dimension $D$.

\subsection{ The solution for $Q_\e^2 = Q_\m^2$}  %
In this degenerate case, solutions can be found which need not satisfy
the orthobrane condition (\ref{4.1}).
Let us suppose that two functions (\ref{3.5}), say, $y_1$ and $y_2$,
coincide
up to an addition of a constant
(which may be then absorbed by re-defining a
charge $Q_1$ or $Q_2$) while
corresponding vectors $Y_1$ and $Y_2$
are neither coinciding, nor orthogonal (otherwise we would have
the previously considered situation). Substituting $y_1\equiv y_2$ into
(\ref{3.16}), one obtains
\beq{4.10}
        (Y_{1,A} - Y_{2,A})x^A =0 .
\eeq
This is a constraint reducing the number of
independent unknowns $x^A$.
Furthermore, substituting (\ref{4.10}) to
the Lagrange equations for $x^A$,
\beq{4.11}
        -(Y_{1,A} - Y_{2,A}){x''}^A =
        \sum_s \theta_s Q_s^2 \e^{2y_s} Y_s^A (Y_{1,A} - Y_{2,A}) =0.
\eeq
In this sum all coefficients of different functions $\e^{2y_s}$
must be zero. This yields new orthogonality conditions
\beq{4.12}
        Y_s^A (Y_{1,A} - Y_{2,A}) =0, \cm s \ne 1,2 ,
\eeq
now for the difference $Y_1-Y_2$ and other $Y_s$, and
with \eq(\ref{4.1a}) the relation
\beq{4.13}
        (\nu^{-1} - Y_1^A Y_{2,A})(\theta_1 Q_1^2 - \theta_2 Q_2^2) =0 .
\eeq
The first multiplier in (\ref{4.13}) is positive ($\oG_{AB}$ is
positive-definite, hence a scalar product of two different vectors with
equal norms is smaller than their norm squared). Therefore
\beq{4.14}
        \theta_1 = \theta_2, \inch Q_1^2 = Q_2^2.
\eeq

Imposing the constraints (\ref{4.10}), (\ref{4.12}), (\ref{4.14}),
reduces the numbers of unknowns and integration constants,
and simultaneously also reduces the number of restrictions on the input
parameters (by the orthogonality conditions (\ref{4.1})).
Due to (\ref{4.14}), this is only
possible when the two components with coinciding charges are of equal
nature:  both must be either true electric/magnetic ones
($\theta_s=1$), or quasiscalar ones ($\theta_s= -1$).
Correspondingly, we now set
$y(u):=y_\e = y_\m $ and $Q^2:=Q_\e^2 = Q_\m^2 $.

For the minimal configuration
(\ref{6.1})--(\ref{6.2a}),
eq. (\ref{4.10}) yields
\beq{6.17}
        d_3\beta_3 -d_4\beta_4 -2\lambda\varphi =0.
\eeq
Eqs.\,(\ref{4.12}) are irrelevant here since we are dealing
with two functions $y_s$ only.
The equations of motion for $x^A$ now take the form
\beq{6.18}
        {x^A}'' = Q^2 \e^{2y} (Y^A_{\e} + Y^A_{\m}).
\eeq
Their proper combination gives $y'' = (1+d_2) Q^2 \e^{2y}$, whence
\beq{6.19}
        \e^{-y} = \sqrt{(1+d_2)Q^2} s (h, u+u_1)
\eeq
where the function $s$ is defined in (\ref{3.10}) and $h,u_1$ are
ICs and, due to (\ref{6.3}),\\ $\sqrt{(1+d_2)Q^2} s (h, u_1)=1$.
Other unknowns are easily determined using (\ref{6.18}) and
(\ref{6.3}):
\bear{6.20}
        x^A \eql \nu Y^A y + c^A;
\cm
                        Y^A = Y^A_{\e} + Y^A_{\m} = (1, 1, 0, 0, -1, 0);
\\\nn
        \sigma_1 \eql -\nu y + c_0 u.
\ear
Here, as in (\ref{6.5a}), $\nu = 1/(1+d_2)$, but it is now just
a notation. The constants $c_0,\ h,\ c^A\ (A=1,\ldots,6)$ and $k$
(see (\ref{3.9})) are related by

\bear{6.21}
&&    -c^0 + \sum_{i=1}^{5} d_i c^i = 0,
\qquad
        c^1 + d_2 c^2 + d_3 c^3 - \lambda c^6 = 0,
\qquad
        c^1 + d_2 c^2 + d_4 c^4 + \lambda c^6 = 0,
\nn\\
&&      2k^2 {\sign k} = {\frac{2h^2 \sign h}{1+ d_2}}
                     (c^0)^2 + \sum_{i=1}^{5} d_i (c^i)^2 + (c^6)^2 .
\ear

Extra-dimensional scale factors
remain finite as $u\to u_{max}$ in the case  of a BH. It is
specified by the following values of the ICs:
\beq{6.22}
     k=h>0,\quad c^3=c^4=c^6=0,\quad
                             c_2=-c_5=-\frac{k}{1+d_2}\ ,
\quad
     c_0=c^1 =-\frac{d_2 k}{1+d_2}\ .
\eeq
The event horizon occurs at $u=\infty$. After the same transformation
(\ref{6.7}) the metric takes the form
\bear{6.23}
     ds^2_D \eql
            - \frac{1-2k/R}{(1+p/R)^{2\nu}}dt^2
     +(1+p/R)^{2\nu}
                  \left(\frac{dR^2}{1-2k/R}+R^2d\Omega^2\right)
\nnnv
\inch\cm
+(1+p/R)^{-2\nu} ds_2^2 +ds_3^2 + ds_4^2 + (1+p/R)^{2\nu}ds_5^2
\ear
with the notation
\beq{6.24}
      p = \sqrt{k^2 + (1+d_2)Q^2}-k.
\eeq
The fields $\varphi$ and $F$ are determined by the relations
\beq{6.25}
     \varphi\equiv 0\ ,
\cm
     F_{01L_3 \ldots L_n} = -\frac{Q} {R^2(1+p/R)},
\cm
     F_{23L_3 \ldots L_n} = Q \sin \theta.
\eeq

\beq{6.26}
G_N M = k +p/(1+d_2),
\eeq
The Hawking temperature can be calculated as before,
\beq{6.27}
     T = \frac{1}{2\pi k_{\rm B}}
             \frac{1}{4k}\left(\frac{2k}{2k+p}\right)^{2/(d_2+1)}.
\eeq The well-known results for the Reissner-Nordstr\"om metric
are recovered when $d_2=0$. In this case $T\to 0$ in the extreme
limit $k\to 0$.  For $d_2=1$, $T$ tends to a finite limit as $k\to
0$ and for $d_2>1$ it tends to infinity. As is the case with two
different charges, $T$ does not depend on the space-time dimension
$D$, but depends on the $p$-brane intersection dimension $d_2$.

\section{\bf The Einstein frame for dynamics and cosmology}
\setcounter{equation}{0}
In this final section we discuss
the issue of the physical frame for the
particularly important case  $D_0=4$.
First of all, in this case a
selfdual canonical
formulation of dynamics
is at hand,
due to the particular spinor decomposition
$\so (1,3)=\su (2)\oplus \su(2)$
of the tangent Lorentz symmetry.
In the Einstein frame,
the effective $\sigma$-model with $D_0=4$
admits in principle
a canonical quantization of the geometry on
$\ol{M}_0=\R\times M_0$
to the same extend and under the same assumptions
as pure Einstein gravity does.

Since for a multidimensional geometry as defined above
the imprint of the internal factor spaces is
only by their scale factors,
configuration space and phase space of such geometries
will only be extended by finite a finite number
of dilatonic midisuperspace fields.
However, only in the Einstein frame the coupling
of the dilatonic fields to the $\ol D_0$-geometry will be minimal
such that the quantization of the latter can be executed
practically independently.

Let us denote the external space-time metric $\ol{g}^{(0)}$
in the Brans-Dicke frame with $\ol{\gamma}\mustbe 0$
as $\overline{g}^{(\BD)}$
and in the Einstein frame with $f\mustbe 0$
as $\overline{g}^{(\E)}$.
It can be easily seen that they are connected with each other
by a conformal transformation
\beq{ef1}
\overline{g}^{(\E)}
\mapsto
\overline{g}^{(\BD)} = \Omega^2 \overline{g}^{(\E)}
\eeq
with $\Omega$ from \rf{Omega}.

In particular, also for spatially homogeneous cosmological models
(and with $t\leftrightarrow u$ for spherically symmetric static models)
solutions have to be transformed to the
Einstein frame before a physical interpretation can be given.

Under any projection $\pr_0: \overline{M_{0}} \to \R$
a consistent pullback
of the metric
$- e^{2\gamma(\tau)} d\tau\otimes d\tau$
{}from $\tau\in \R$ to $x\in\pr^{-1}_0\{\tau\}\subset \ol{M}_0$
is given by
\beq{ef3}
\overline{g}^{(\BD)}(x)
:= - e^{2\gamma(\tau)} d\tau\otimes d\tau
+ e^{2\beta^0(x)} {g}^{(0)} .
\eeq
In particular,
for spatially (metrically-)homogeneous cosmological models
all scale factors $a_i:=e^{\beta^i}$, $i=0,\ldots,n$,
depend only on $\tau\in\R$.

With \rf{ef3} and \rf{ef1}, Eq.
\rf{1.2} reads
\bear{ef4}
g
&=& - e^{2\gamma(\tau)} d\tau\otimes d\tau + a_{0}^2 g^{(0)}
+\sum_{i=1}^{n} e^{2\beta^i} g^{(i)}
\nn\\
&=& -  dt_{\BD} \otimes dt_{\BD} + a^2_{\BD} g^{(0)}
+\sum_{i=1}^{n} e^{2\beta^i} g^{(i)}
\nn\\
&=&  -  \Omega^2 dt_{\E} \otimes dt_{\E} + \Omega^2 a^2_{\E} g^{(0)}
+\sum_{i=1}^{n} e^{2\beta^i} g^{(i)} ,
\ear
where $a_0:= a_{\BD}$ and $a_{\E}$ are the external space scale factor
functions depending respectively on the cosmic synchronous time
$t_{\BD}$ and $t_{\E}$ in the Brans-Dicke and Einstein frame.
With \rf{Omega} the latter is related to the former by
\beq{ef5}
a_{\E}
= \Omega^{-1} a_{\BD}
=\left(
\prod_{i=1}^n e^{d_i\beta^i}
\right)^{ \frac{1}{{D}_0-2} }
a_{\BD} ,
\eeq
and the cosmic time of the Einstein frame is given by
\beq{ef6}
\pm dt_{\E}
=\Omega^{-1}e^{\gamma}d\tau
=
\left(
\prod_{i=1}^n e^{d_i\beta^i}
\right)^{ \frac{1}{{D}_0-2} }
dt_{\BD}
\ . \eeq Since $a^2_{\BD}(d\eta_{\BD})^2=\Omega^{2}a^2_{\E}
(d\eta_{\E})^2$, the conformal times of the Einstein and the
Brans-Dicke frame agree (up to time reversal). This has sometimes
guided authors to compare the frames in conformal time (see e.g.
\cite{GV}). However (at least for cosmology) the physical relevant
time is the cosmic synchronous time, which is different for
different frames, in particular for the Einstein and Brans-Dicke
frame.

In \cite{RZ98} several reasons have been listed why minimal
coupling between geometry and matter and hence the Einstein frame
is the preferred choice. There also a general prescription for the
transformation of well known solutions from the Brans-Dicke frame
to the Einstein frame has been given. It was demonstrated
explicitly that qualitative cosmological features change
significantly under this transformation. This was shown for a
couple of examples, including the general multidimensional Kasner
solution and a special inflationary solution with constant
internal volume. In particular it was shown that inflationary
solutions in the Brans-Dicke frame transform into non-inflationary
ones in the Einstein frame. It is to be expected that this is a
rather general feature, whence the multitude of solutions which
appear inflationary  in the Brans-Dicke frame will be indeed
non-inflationary when considered in the Einstein frame.

\section{\bf Discussion}
\setcounter{equation}{0}
The Einstein action with minimally coupled scalars and $p$-branes
in higher dimension $D$ can be reduced to an effective model in
lower dimension ${D}_0$. This results in a (generalized)
$\sigma$-model with conformally flat target space. With a purely
geometrical dilaton field $f$, it provides a natural
generalization for the well-known Brans-Dicke theory.

The orthobrane condition \rf{i3.19} allows us to find exact
solutions. Furthermore, the orthobrane solution is a sufficient
condition for the target space of the $\sigma$-model to be a
locally symmetric space. The orthobrane case is the generic one
(apart from cases with degenerate coupling matrix  \rf{Lmat})
where the target space is locally symmetric.

Examples of a certain minimal static, spherically symmetric
$p$-brane configuration are given with just one electric and one
magnetic antisymmetric $F$ component (since in 4 dimensions we
only deal with a single electromagnetic field), which in general
intersect and interact with a single scalar field. Spherical
symmetry here is considered in the physical relevant $D_0=4$ case
of $S^2$ spheres, although the extension to arbitrary spheres is
straightforward.

Besides popular families of orthobrane solutions there are further
families of solutions, which have another additional symmetry,
e.g. coinciding $F$-field charges for the electro-magnetic
solutions. In the target space this additional symmetry is
expressed by a linear dependency \rf{4.10} between column vectors
$Y$ of the coupling matrix $L$ defined in \rf{Lmat}.

For the mentioned static solutions, Hawking temperature $T_H$ can
be formally calculated by surface gravity via a Komar-like
integral. For both, the orthobrane case and the case of equal
charges $Q^2_e=Q^2_m$, the expressions of $T_H$ depend
characteristically on the intersection dimension.
This results are also interesting in the context of  recent
increased interest in extremal $p$-brane configurations with black
holes \cite{Sat}.

The  interpretation of the extremal limit $k\to 0$ is delicate.
The solutions above have been described in isotropic coordinates
which cover just  the asymptotically flat exterior of the black
hole. A better understanding of their global causality  structure
would require an investigation of the maximal extension of the
space-time rather than only of its exterior part. The limit $k\to
0$ was here called extremal, since via \rf{6.13} and \rf{6.26} in
this limit the effective asymptotical Schwarzschild mass $M$ is
just given by the charges, $G_N M= \Half (B p_\e + C p_\m) $ and $
G_N M =p/(1+d_2)$, respectively. Further work is required to
understand this type of extremality, and the related  asymptotics
of $T_H$, which remains finite for intersection dimension $d_2=1$
and becomes infinite for $d_2\geq 2$. As it was pointed out
recently in \cite{ABCK} particular care is needed in order to
associate the correct physical charges and thermal properties of a
black hole correctly with its horizon.

The multidimensional $\sigma$-model opens the door for further
investigations, in particular also for covariant and canonical
quantization. The effective $\sigma$-model reduction appears as a
possible clue to canonical quantization within a large well
defined class of higher dimensional geometries, namely the
multidimensional ones. Covariant quantization techniques can be
applied in any dimension $D_0$. In particular, they are well
applicable to our new solutions with scalar fields only, when the
target space is flat. The effective geometry of the
$D_0$-dimensional model can be reformulated in terms of
connections and  soldering forms. For $D_0=4$, not only a
canonical $1+3$ split can be performed, but moreover the canonical
quantization of the $D_0$-geometry can be performed with self-dual
variables in the usual manner.

Finally, in analogy to investigations in \cite{GalR}, it should be
possible to apply solution generating techniques like the
Ehlers-Harrison transformation also in the context of the
multidimensional $\sigma$-model.

\nl\nl
{\Large {\bf Acknowledgments}}
\nl\nl
I would like to thank to
A. Ashtekar, O. Brodbeck, K. Bronnikov, S. Das, L. Freidel, S. Gupta,
V. Ivashchuk, J. Pullin, R. Puzio, and L. Smolin for discussions
 on the subject, and to N. Guerras for permanent encouragement.
This work was supported by NSF grant PHY-9514240 to The
Pennsylvania State University and a gift from the Jesse Phillips
Foundation.

\np\noindent

\end{document}